\documentclass[aps,twocolumn]{revtex4}
\usepackage{epsfig}
\usepackage{graphicx}
\usepackage{amsmath,amssymb,amsfonts}
\usepackage{array}
\usepackage{url}
\usepackage{multirow}
\usepackage{subcaption} 
\usepackage{amsmath}
\usepackage{booktabs}
\usepackage{float}
\usepackage{lineno}
\usepackage{xspace}
\usepackage{ulem}
\usepackage{caption}
\usepackage{algorithm}
\usepackage{placeins} 
\usepackage{algpseudocode}
\usepackage{tikz}
\usepackage{xcolor}
\usetikzlibrary{fit, positioning}
\definecolor{darkblue}{RGB}{0,0,196}
\usepackage[colorlinks=true,linkcolor=darkblue,citecolor=darkblue,urlcolor=darkblue]{hyperref}

\newcommand{\pT}{p_{\mathrm{T}}}
\newcommand{\Nch}{N_{\mathrm{ch}}}
\begin{document}

\setlength{\heavyrulewidth}{0.02em}
\setlength{\lightrulewidth}{0.01em}
\title{Inferring identified hadron production in $pp$ collisions \\with physics-informed machine learning at the LHC}

\author{Rishabh Gupta$^{1}$}
\author{Kangkan Goswami$^{1}$}
\author{Suraj Prasad$^{1,2}$}
\author{Raghunath Sahoo$^{1}$}\email[Corresponding Author: ]{Raghunath.Sahoo@cern.ch}

\affiliation{$^{1}$Department of Physics, Indian Institute of Technology Indore, Simrol, Indore 453552, India}
\affiliation{$^2$HUN-REN Wigner Research Centre for Physics, 29-33 Konkoly-Thege Miklós Str., H-1121 Budapest, Hungary}
\begin{abstract}
Machine learning has become a powerful tool in high-energy collider experiments, which enables the studies based on data-driven approaches to complex reconstruction and regression tasks. The study of identified hadron spectra in pseudorapidity regions beyond detector acceptance, which is limited to mid-rapidity regions, carries important information about particle production, yet remains unmeasured. In this work, we develop a physics-informed neural network, trained on PYTHIA8 $pp$ collisions at $\sqrt{s}=13.6$~TeV, to infer $\pT$ spectra of $\pi^{\pm}$, $K^{\pm}$, $p/\bar{p}$, $\Lambda/\bar{\Lambda}$, and $K^{0}_{\mathrm{s}}$ in different rapidity regions. Physics-motivated constraints, including particle yield ratios, spectral shape, and smoothness, are incorporated into the loss function. A staged hyperparameter optimization strategy is used to ensure stability. The model achieves yield uncertainties of ${\sim}1.5\%$, $1.8\%$, and $5.83\%$ in the training, interpolation, and extrapolation regimes, respectively, outperforming XGBoost and LightGBM. It further reproduces key observables such as particle yield ratios, the multiplicity dependence of $\langle \pT \rangle$, and kinetic freeze-out parameters, indicating that the model captures the underlying physics and provides reliable predictions beyond the measured phase space.
\end{abstract}

\maketitle

\section{Introduction}
To understand the fundamental properties of strongly interacting matter, ultra-relativistic proton-proton and heavy-ion collisions are studied at accelerator facilities such as the Relativistic Heavy-Ion Collider (RHIC) at BNL and the Large Hadron Collider (LHC) at CERN. These collisions provide access to matter at extremely high temperature and energy density, where quarks and gluons form a deconfined state of matter known as the Quark Gluon Plasma (QGP). However, due to its short lifetime, the QGP phase can not be observed directly, and only the final-state hadrons are experimentally detected. These final-state hadrons carry important information about particle production mechanisms, system evolution, and hadronization dynamics. Therefore, the study of identified particle production in high-energy collisions serves as an important tool to understand the QGP ~\cite{McLerran:1986zb, Wong:1995jf, Gyulassy:2004zy,Shuryak:2014zxa}. 

Particle production in high-energy hadronic and heavy-ion collisions shows a strong dependence on both transverse momentum ($\pT$) and pseudorapidity ($\eta$), reflecting different underlying QCD mechanisms. At low $\pT$, particle yields are dominated by soft processes including but not limited to string fragmentation and Multi-Parton Interactions (MPI) with a bulk production of particles, leading to an approximately exponential spectral shape and clear mass-dependent effects. The mass-dependent effects are usually related to soft or thermal productions and are therefore crucial in the understanding of underlying production mechanisms. In contrast, the high-$\pT$ region is governed by hard partonic scatterings, where perturbative QCD becomes applicable, and the particle spectra follow a power-law behavior. The dependence on $\eta$ encodes the longitudinal dynamics: particle production is highest around midrapidity ($\eta \approx 0$), where particle densities are largest, and decreases toward forward regions due to kinematic constraints and parton distribution effects~\cite{ALICE:2020nkc}. Together, the $(\pT, \eta)$ dependence provides a comprehensive picture of the interplay between soft and hard QCD processes in multi-hadron production dynamics. 

Moreover, an accurate reconstruction of particle spectra is essential for the extraction of bulk properties of the produced system, such as kinetic freeze-out temperature and radial flow velocity, which are commonly obtained through model fits to the spectra~\cite{Schnedermann:1993ws, Prasad:2021bdq, ALICE:2013mez, MenonKavumpadikkalRadhakrishnan:2023cik}. However, the experimental measurements are limited by detector acceptance, which results in restricted pseudorapidity coverage~\cite{ALICE:2008ngc, ALICE:2023udb}. This limits the ability to understand the particle production in these pseudorapidity gaps. This limitation motivates the development of data-driven approaches to reconstruct particle yields in unmeasured $\eta$ regions.

In high-energy physics, Machine Learning (ML) methods have been successfully applied to a wide range of problems, including reconstruction, classification, and regression tasks~\cite{Sahoo:2025vip, Goswami:2024xrx, Prasad:2023zdd, Mallick:2021wop, Mallick:2023vgi, Mallick:2022alr, Tiwari:2026vef}. The ML techniques provide a reliable framework, due to their ability to model high-dimensional, non-linear dependencies without imposing an explicit functional form. Therefore, the ML models can learn the dependence of yields on $\pT$, $\eta$, and multiplicity (or centrality), which enables us to predict the identified hadron yield in regions where direct measurements are not available. Moreover, the interpolation within the training range is generally well controlled, but the extrapolation to unmeasured $\eta$ regions is significantly more challenging due to the absence of direct constraints from experimental data. In such cases, the model must rely on learned correlations and underlying physical structure to produce reliable predictions. In some cases, the machine learning models may reproduce the particle yield in the training range accurately, but fail to preserve physically relevant features in the extrapolation region. Therefore, ensuring that the model respects known physical properties of particle spectra is essential for reliable predictions. To address this challenge, it is necessary to incorporate physics-motivated constraints directly into the learning framework. This can be achieved within the framework of physics-informed neural networks (PINNs), where domain-specific physical constraints are embedded directly into the training objective, thereby guiding the model toward physically consistent solutions even in regions where data are sparse or unavailable. Such constraints can guide the model toward physically consistent solutions by enforcing smooth spectral behavior, preserving relative particle yields, and maintaining the structural properties of the spectra across kinematic regions. This enables the model to generalize beyond the training range and follow the known physical principles.

In this work, we predict identified hadron yields for different $\eta$ regions using machine learning regression models trained on Monte Carlo events generated with PYTHIA8. The models are designed to learn the multi-dimensional dependence of the spectra on $\pT$, $\eta$, and multiplicity. Moreover, to ensure physically consistent predictions, we incorporate physics-informed constraints directly into the training objective, enforcing spectral smoothness, shape consistency, and particle yield ratios. Moreover, we ensure stable optimization in the presence of data-driven and physics-based objectives by employing a staged hyperparameter optimization (HPO) strategy. The model is first trained to capture the data-driven structure, and it is subsequently refined under progressively stronger physics constraints. Furthermore, we explicitly evaluate the model performance in both interpolation and extrapolation regions. This provides a controlled framework to assess the ability of physics-informed machine learning models to reconstruct particle spectra in unmeasured regions.

The rest of the paper is organized as follows. Section~\ref{sec:methodology} discusses the data generation and ML model architecture, and model training. In Sec.~\ref{sec:results}, we compare different ML models in different kinematic regions. Some physics observables sensitive to particle $\pT$ spectra are predicted and compared with their true values. Finally, the paper is summarized in Sec.~\ref{sec:summary} with a brief outlook.

\section{Methodology}
\label{sec:methodology}
\subsection{Dataset Generation and Pre-processing}

\subsubsection{Dataset Generation}
As the primary objective of this work is to study particle yield distributions as a function of $\pT$, $\eta$, and $N_{\rm ch}$, we generate minimum-bias $pp$ collisions at $\sqrt{s}=13.6$ TeV using PYTHIA8 within the MAGE pipeline~\cite{mage_ref}. PYTHIA8 is a widely used Monte Carlo event generator designed to simulate leptonic, hadronic, and heavy-ion collisions over a broad range of center-of-mass energies, from GeV to TeV scales. It models the complete evolution of a collision event, beginning with the initial hard scattering of partons, followed by parton showers, multiparton interactions, and culminating in the hadronization process \cite{Sjostrand:2014zea,Sjostrand:2006za,Prasad:2023zdd,Prasad:2025yfj,Prasad:2024gqq,Goswami:2024xrx,ALICE:2023bga,ALICE:2015ikl,Thakur:2017kpv,Deb:2018qsl,Bierlich:2026syi}. MAGE is a reproducibility-first workflow framework designed to simplify large-scale simulation and analysis tasks. It encapsulates the workflow configuration, execution settings, and output-handling steps within project files carrying the \texttt{.mgp} extension, allowing the same setup to be reproduced and executed across multiple runs with minimal manual intervention. This also facilitates parallel execution while reducing the need for users to directly manage software version dependencies or handle ROOT-based output. In the present analysis, the MAGE pipeline serves as the interface between event generation and downstream data processing, enabling conversion of the generated ROOT output~\cite{Brun:1997pa} directly into user-defined formats. For the PYTHIA8 simulations, we employ the Monash tune~\cite{Skands:2014pea}, which provides a reasonable description of experimental data from $pp$ collisions at LHC energies~\cite{Prasad:2024gqq,Prasad:2025yfj}.

Particle yields for $\pi^{\pm}$, $K^{\pm}$, $p/\bar{p}$, $\Lambda/\bar{\Lambda}$, and $K^{0}_{\mathrm{s}}$ are obtained from approximately $10^{10}$ minimum bias events for pp collisions at $\sqrt{s}=13.6$ TeV in the pseudorapidity range $0 \leq |\eta| < 3$. The data are binned with a pseudorapidity interval of $\Delta\eta = 0.1$ and a charged-particle multiplicity bin width of $\Delta\Nch = 10$. The dataset is subsequently passed to the preprocessing stage.

\subsubsection{Dataset Preprocessing and Feature Selection}
To evaluate the applicability of machine learning models across different $\eta$ ranges, we construct a spectral representation that captures the dependence on track and event-level variables. The generated dataset is transformed into $\pT$ spectra for different $N_{\rm ch}$, PID, and $\eta$ ranges. For this study, we consider three multiplicity classes, class~1 ($0 < N_{\rm ch} < 20$),  class~2 ($20 \leq N_{\rm ch} < 50$), and class~3 ($N_{\rm ch} \geq 50$). These multiplicity classes allow us to work with reasonable statistics in all $\pT$ bins up to 10 GeV/\textit{c}. This selection of $N_{\rm ch}$ classes also ensures that the classification between the high and low multiplicity events is reasonable to account for the underlying event activity.

The input feature vector for each data point consists of the $\pT$ bin index, the multiplicity class index (encoded as \{0, 1, 2\}), the PID, and the $\eta$ bin index. Here, the PID labels carry no inherent numerical ordering for individual particles, and hence, we encode them as one-hot vectors, allowing the model to treat each particle species as a distinct category. Moreover, to work with better statistics, we combine the $\eta$ bins such that we have $0 \leq |\eta| < 0.2$, $ 0.2 \leq |\eta| < 0.4$ up to $ 2.8 \leq |\eta| < 3.0$. Furthermore, the $\eta$-bin index is assigned sequentially, so that $0 \leq |\eta| < 0.2$ corresponds to index 0, $ 0.2 \leq |\eta| < 0.4$ to index 1, and so on, which gives 15 indices over the entire generated range. The binning adopted in this study is given in Table~\ref{tab:dataset_summary}.

To improve the model's sensitivity to all input features, the feature variables are standardized using $z$-score normalization, denoted by $\mathcal{Z}$. This transformation rescales each feature to have zero mean and unit variance, improving the stability and convergence of neural network training because models are sensitive to the relative scale of input variables. The transformation is defined as
\begin{equation}
\label{eqn:z_scoring}
\tilde{x}_{n,j} = \mathcal{Z}(x_{n,j})
= \frac{x_{n,j} - \mu_j}{\sigma_j}
\end{equation}
where $\tilde{x}_{n,j}$ represents the transformed value of feature $j$ for the $n^{\text{th}}$ data point, and $\mu_j$ and $\sigma_j$ are its mean and standard deviation, respectively, computed over the training set as,
\begin{equation}
\label{eqn:mean_z_scoring}
\mu_j = \frac{1}{N} \sum_{i=1}^{N} X_{i,j},
\qquad
\sigma_j = \sqrt{\frac{1}{N} \sum_{i=1}^{N} \left(X_{i,j} - \mu_j\right)^2}
\end{equation}
Here, $N$ is the number of training data points, and $X_{i,j}$ indicates the value of the $j^{th}$ feature for the $i^{th}$ data point.

The final dataset consists of individual data points extracted from the projected spectra, each with 8 input features (indices for $\pT$ bin, $\eta$ bin, multiplicity class, and 5 one-hot encoded PID) and one target output, as defined in Eqs.~\eqref{eqn:input_feature} and ~\eqref{eqn:output_target}
\begin{equation}
\label{eqn:input_feature}
\tilde{\mathbf{x}}_n =
\left(
\mathcal{Z}(\pT^{(i)}),\;
\mathcal{Z}(k),\;
\mathcal{Z}(c^{(j)}),\;
\mathbf{e}_{\mathrm{PID}}
\right).
\end{equation}
Here, $\pT^{(i)}$ denotes the bin value corresponding to the $i^{\text{th}}$ index, $k$ represents the $\eta$-window bin index, $c^{(j)}$ denotes the $N_{\rm ch}$ class for the $j^{\text(th)}$ spectrum, and $\mathbf{e}_{\mathrm{PID}}$ is a one-hot vector that is equal to 1 for the corresponding particle species and 0 for all others.

To mitigate the dominance of large values in the loss function and ensure the sensitivity to low-yield regions, the model is trained to predict the target on a log scale. The target output is given as,
\begin{equation}
\label{eqn:output_target}
\begin{aligned}
y_n = \log \Bigg[
\frac{1}{N_{\mathrm{events}}}
\frac{d^2 N_{(\mathrm{PID},\,\eta_k)}}{d\pT \, d\eta}
\left(\pT^i\right)
\Bigg]
\end{aligned}
\end{equation}

The datapoints constructed from the dataset follow the structure given as,
\begin{equation}
    \label{eqn:datapoints}
    \mathcal{D} = \left\{ \left( \tilde{\mathbf{x}}_n,\,y_n\right) \right\}_{n=1}^{N}
\end{equation}
where $n$ runs from 1 through the total number of data points. A complete summary after preprocessing is given in Table~\ref{tab:dataset_summary}.

\begin{table}[h]
\centering
\caption{Summary of the dataset and selection regions.}
\label{tab:dataset_summary}
\begin{tabular}{lc}
\toprule
Property & Value \\
\midrule
Collision system & $pp$, $\sqrt{s} = 13.6$~TeV \\
Total events & $9.7\times10^{9}$ \\
$\eta$ range & $0 \leq |\eta| < 3$ \\
$\eta$-bin width & 0.4 \\
$\eta$ bins & 15 \\
Multiplicity classes & 3 \\
Particle species & $\pi^{\pm}$, $K^{\pm}$, $p/\bar{p}$, $\Lambda/\bar{\Lambda}$, $K^0_s$ \\
$\pT$ range & $0.0$--$10.0$~GeV/$c$ \\
$\pT$ bins & 40 (non-uniform) \\
$\pT$ binning & 
$\begin{aligned}
&0.0 \le \pT < 3.0:\ \Delta \pT = 0.1 \\
&3.0 \le \pT < 6.0:\ \Delta \pT = 0.5 \\
&6.0 \le \pT \le 10.0:\ \Delta \pT = 1.0
\end{aligned}$ \\
Total spectra & 225 \\
Total datapoints & 9000 \\
Input features per datapoint & 8 \\
\midrule
\textbf{Dataset splits (in $|\eta|$)} & \\
Training region & $|\eta| < 1.2$, $1.4 \le |\eta| < 1.6$ \\
Interpolation region & $1.2 \le |\eta| < 1.4$ \\
Extrapolation region & $1.6 \le |\eta| < 3.0$ \\
\midrule
Training/Validation split&  90/10\\
\bottomrule
\end{tabular}
\end{table}

\subsection{Model Architectures}
The primary objective of this study is to predict transverse-momentum spectra in different pseudorapidity regions. We also evaluate whether machine learning models can capture the evolution of spectral shapes across $\eta$ while preserving the underlying physics of particle production. We compare three classes of models, gradient boosted decision trees (XGBoost~\cite{ml:xgboost} and LightGBM~\cite{ml:2017lightgbm}) and neural networks, which have demonstrated good performance in regression tasks and related high-energy physics applications~\cite{Mallick:2023vgi, Mallick:2022alr, Prasad:2023zdd, Goswami:2024xrx, Shokr:2021ouh}. Moreover, to enforce physically meaningful behavior in the neural network, we employ a physics-informed neural network (PINN) framework.  In this work, the PINN is implemented through additional loss terms that enforce spectral smoothness, shape consistency, and particle yield ratios. These constraints guide the model toward physically consistent solutions, particularly in extrapolation regions where direct training data are not available. The model architectures considered in this study are described below.

\subsubsection{XGBoost}
XGBoost, short for eXtreme Gradient Boosting, is an advanced implementation of gradient-boosting decision trees (GBDT) that incorporates several optimizations to improve speed and accuracy. It is widely used due to its flexibility across a range of tasks, including classification, regression, and ranking.
XGBoost formulates gradient boosting using a second-order Taylor expansion of the loss function, enabling a more accurate approximation of the optimization objective~\cite{ml:xgboost}.  Further, it employs parallel processing by splitting candidates to evaluate different features simultaneously, reducing training time. In addition, it uses tree pruning, a regularization technique that removes splits that do not provide sufficient improvement in the objective function, thereby preventing overfitting and producing more generalizable models. Furthermore, XGBoost offers a wide range of hyperparameters that can be tuned to optimize model performance \cite{ml:xgboost_docs}.

\subsubsection{LightGBM}
LightGBM (Light Gradient Boosting Machine) is an efficient implementation of gradient-boosted decision trees designed to reduce computational cost and memory usage compared to conventional boosting methods~\cite{ml:2017lightgbm}. In contrast to the traditional level-wise tree growth strategy, where all nodes at a given depth are expanded uniformly, LightGBM adopts a leaf-wise growth strategy, where at each step the leaf with the maximum reduction in loss is selected for splitting. This produces an asymmetric tree that focuses on the most informative regions of the feature space, typically achieving faster convergence and higher accuracy on large-scale datasets. The implementation of LightGBM is described in Ref.~\cite{ml:lightgbm_docs}.

\subsubsection{Physics-Informed Neural Network}
Physics-informed neural networks are a class of neural networks trained to respect laws of physics described by general nonlinear partial differential equations. Instead of solely relying on data, PINNs enforce governing equations as constraints, typically expressed as differential equations, to guide the model towards physically consistent solutions. This embedding of domain knowledge improves generalization in sparse data regimes and reduces dependence on large training datasets. Furthermore, it ensures that predictions remain consistent with established theoretical frameworks, even beyond the training distribution. In this work, we employ a generalized PINN approach, in which domain-specific physical constraints are incorporated into the training objective instead of enforcing differential operators, as in the original formulation~\cite{Raissi:2019pinn,Karniadakis:2021}.

The PINN for predicting particle spectra requires constraints that enable the model to capture both point-wise values and the underlying structure of spectra across different kinematic regions and particle species, without over-constraining the optimization. To achieve this, we combine a data-driven loss with additional structure-aware regularization terms. The primary objective is defined as a weighted data loss,
\begin{equation}
    \mathcal{L}_{\mathrm{data}} = \frac{1}{N} \sum_{i} w_i \, \mathcal{L}(\hat{y}_i, y_i),
\end{equation}

where $\mathcal{L}$ denotes a base loss function (Mean Absolute Error (MAE), Huber, or Mean Squared Error (MSE)), selected through staged evaluation. The weights are defined as

\begin{equation}
    w_i = \min\!\left[
\frac{1}{\max\!\left[(\sigma_i / Y_i)^2,\ \epsilon\right]},\
w_{\max}
\right]
\end{equation}

where, $\sigma_i$ is the statistical uncertainty on the yield $Y_i$ and 

$\varepsilon = 10^{-4}$ to prevent divergence when $\sigma_i \to 0$, and $w_{\max}=10$ to prevent any single bin from dominating the loss.

To enforce smoothness, we introduce a curvature regularization term based on second-order finite differences,

\begin{equation}
    \Delta^2 \hat{y}_i = \hat{y}_{i+1} - 2\hat{y}_i + \hat{y}_{i-1},
\end{equation}
\begin{equation}
    \mathcal{L}_{\mathrm{smooth}} = \left\langle \left( \Delta^2 \hat{y}_i \right)^2 \right\rangle.
\end{equation}
Since, $\pT$ ranges are discrete with spacing $\Delta \pT$, this term approximates the second derivative as
\begin{equation}
    \Delta^2 \hat{y}_i \approx \frac{d^2 \hat{y}}{d\pT^2}\bigg|_{p_{{\rm T},i}} (\Delta \pT)^2,
\end{equation}
suppressing rapid changes in slope and enforcing smooth, physically consistent spectral behaviour.

We additionally introduce a shape loss that enforces consistency of local variations between predicted and target spectra,
\begin{equation}
    \mathcal{L}_{\mathrm{shape}} = \left\langle \left( \Delta \hat{y}_i - \Delta y_i \right)^2 \right\rangle,
\end{equation}
where $\Delta y_i = y_{i+1} - y_i$ and $\Delta \hat{y}_i = \hat{y}_{i+1} - \hat{y}_i$. For a fixed $\pT$ range, this is proportional to the squared difference in local slopes between prediction and target,
\begin{equation}
    \left( \Delta \hat{y}_i - \Delta y_i \right)^2 =
    \left( \frac{d\hat{y}}{d\pT} - \frac{dy}{d\pT} \right)^2 (\Delta \pT)^2.
\end{equation}

To preserve physically meaningful relationships between different particle species, we introduce a particle ratio-based constraint in the loss function. Since the particle yields are represented in logarithmic form during training, the differences between the log-yields correspond directly to logarithms of yield ratios, i.e.,
\begin{equation}
    \mathcal{L}_{\mathrm{ratio}} =
    \left\langle \left[ (\hat{y}_a - \hat{y}_b) - (y_a - y_b) \right]^2 \right\rangle .
\end{equation}
This enforces consistency of particle yield ratios across the pairs ($K/\pi$), ($p/\pi$), and ($\Lambda/K^0_S$), which highlights important physical features in the spectra arising from the consolidated effects of strangeness production, baryon number conservation, and radial flow.

Finally, the total loss is given by,
\begin{equation}
  \mathcal{L}_{\rm total} =
  \mathcal{L}_{\rm data}
  + \lambda_r\,\mathcal{L}_{\rm ratio}
  + \lambda_s\,\mathcal{L}_{\rm smooth}
  + \lambda_g\,\mathcal{L}_{\rm shape},
  \label{eq:loss_total}
\end{equation}

The relative contributions are controlled by the weight coefficients $\lambda_r$, $\lambda_s$, and $\lambda_g$, which are optimized using the staged training strategy described in the following section.

Physics-informed loss functions are used in every $\tau = 5$ training epochs. This reduces computational costs while ensuring each physics update acts on a sufficient sample range for the model to learn the structure of the training distribution. In the intermediate epochs, only $\mathcal{L}_{\rm data}$ contributes to the gradient update via standard mini-batch stochastic gradient descent. In every physics epoch, the physics regularization is applied as a separate gradient update following the data loss step,
\begin{equation}
  \theta \;\leftarrow\; \theta
  \;-\; \alpha\,\nabla_\theta\,\mathcal{L}_{\rm data},
  \text{\hspace{0.2cm}}
  \theta \;\leftarrow\; \theta
  \;-\; \alpha\,\nabla_\theta\,\mathcal{L}_{\rm phys},
  \label{eq:two_pass}
\end{equation}

where, $\mathcal{L}_{\rm phys} = \lambda_r\mathcal{L}_{\rm ratio}
+ \lambda_s\mathcal{L}_{\rm smooth}
+ \lambda_g\mathcal{L}_{\rm shape}$.

This decoupled update scheme follows the alternating minimization principle~\cite{Bezdek2003}. At each data epoch, the data loss is minimized across all mini-batches using standard stochastic gradient descent. At every $\tau$ epochs, an additional gradient update based on the physics-informed loss is applied, acting as a periodic correction that guides the model toward physically consistent solutions. The weighting coefficients $\lambda_r$, $\lambda_s$, and $\lambda_g$ are kept small relative to the data loss scale, ensuring the physics correction does not destabilize the optimization trajectory established by $\mathcal{L}_{\rm data}$. The two-step procedure in Eq.~\eqref{eq:two_pass} provides an efficient approximation to joint minimization of $\mathcal{L}_{\rm total}$ in Eq.~\eqref{eq:loss_total}, with the physics update guiding the model toward physically consistent solutions after each data-driven gradient step.

\subsection{Model Selection and Training}
\label{sec:model_selection}
In addition to designing the appropriate model architecture, identifying the optimal model and its hyperparameters is crucial for accurate prediction of $\pT$ spectra. It also ensures that the model effectively captures the underlying physics features. Generally, model hyperparameters, including network architecture and optimization parameters such as \texttt{n\_estimators} in the case of tree-based models, are tuned using a Bayesian optimization method called the Tree-structured Parzen Estimator (TPE)~\cite{Watanabe:2023tpe} via the Optuna framework~\cite{Akiba:2019optuna}. 

\begin{figure*}[t]
    \centering
    \includegraphics[
        width=\textwidth,
        height=0.43\textheight,
        keepaspectratio
    ]{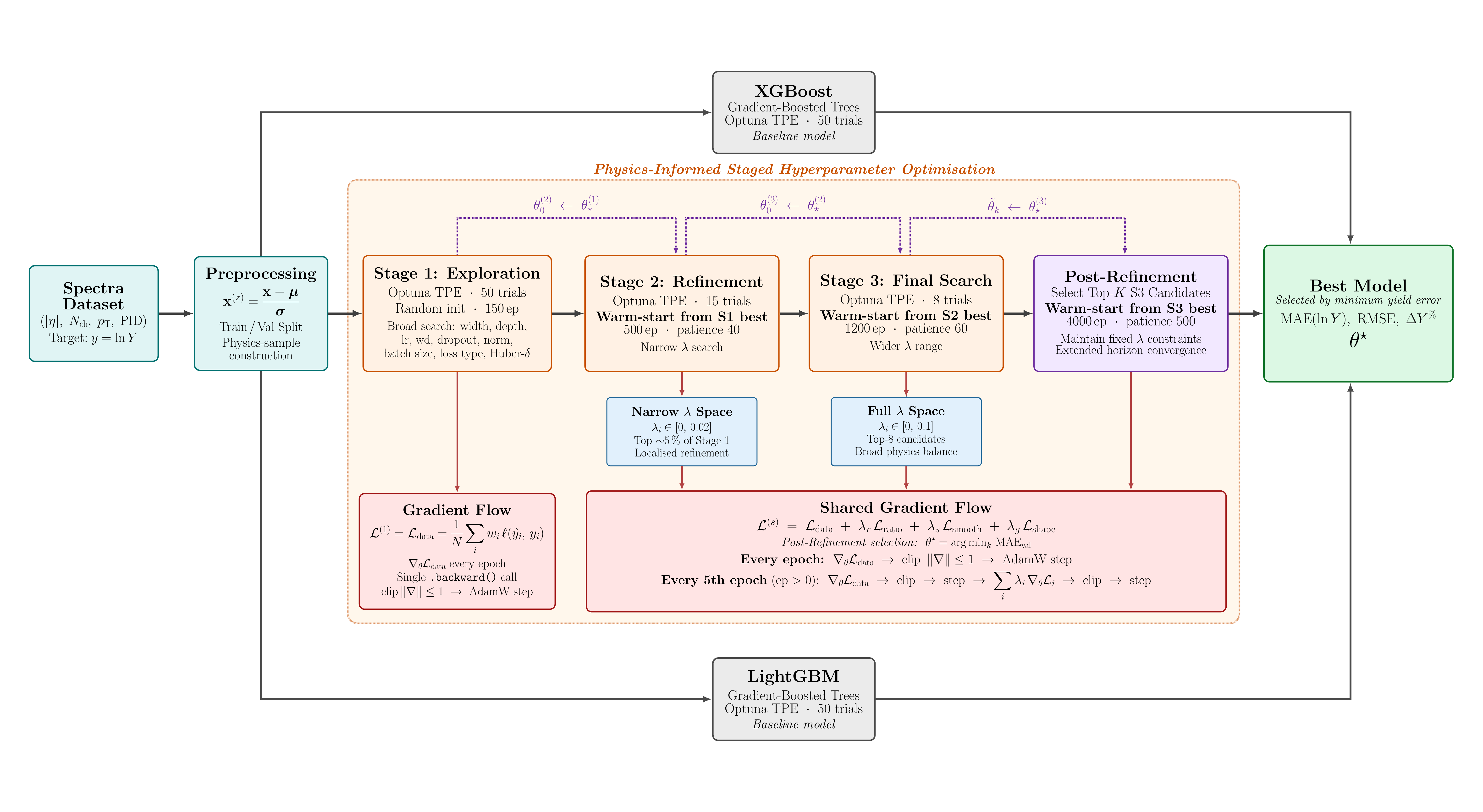}
    \caption{Machine learning pipeline including dataset preparation, staged optimization, and physics-informed neural network training.}
    \label{fig:ml_pipeline_full}
\end{figure*}

The model selection and hyperparameter optimization procedures differ between neural networks and tree-based models. For the PINNs with physics-based loss functions, the relative weights of the loss components are also treated as hyperparameters. To ensure stable and efficient training, we employ a physics-informed staged hyperparameter optimization strategy in which physics-based constraints are progressively incorporated during training, guiding the model toward physically consistent solutions while maintaining predictive performance.

Hyperparameters for XGBoost and LightGBM are optimized with Optuna using TPE sampling across 50 trials each, minimizing validation MAE on log-yield predictions. The loss function is included in the search space as a categorical variable over \{MSE, MAE, Huber\}. When Huber loss is selected, the threshold parameter $\delta$ is additionally sampled as a conditional hyperparameter. No staged strategy is applied to the tree-based models; all hyperparameters are optimized in a single search phase, after which the model is fully trained using the best hyperparameter configuration identified by the search.
The best-performing configuration from each model is selected based on the yield error, defined as

\begin{equation}
    \label{eqn:yield_error}
    \varepsilon_{\mathrm{yield}} =
    \left(e^{\mathrm{MAE}} - 1\right) \times 100\%,
\end{equation}

where the MAE is computed in log-yield space as

\begin{equation}
    \mathrm{MAE} = \frac{1}{N}\sum_{i}
    \left|\log y_i^{\mathrm{pred}}
    - \log y_i^{\mathrm{true}}\right|.
\end{equation}
The physics-informed staged hyperparameter optimization for the PINN is described in detail below. The complete model selection workflow
for all three models is illustrated in Fig.~\ref{fig:ml_pipeline_full}.

\subsubsection{Physics-Informed Staged Hyperparameter Optimization}
Optimizing the model over the full hyperparameter search space, along with physics-informed loss functions, is computationally expensive and, across many trials, can lead to non-physical predictions. After experimenting with different optimization strategies, we find that the HPO workflow for the PINN can be broken down into three stages: a fixed percentage of the top-performing models, evaluated on the validation set, yield error in Eq.~\eqref{eqn:yield_error}, and are promoted to the next stage. The proposed HPO strategy shares structural similarities with FastBO~\cite{Jiang:2024} and BOHB~\cite{Stefan:2018}. However, our approach is specifically tailored to the present problem, as it enables a decoupled gradient update scheme that prioritizes exploring and validating physics-informed loss functions, rather than model architecture, which is effectively constrained during the early stage of optimization. The staged workflow is described below.
\begin{itemize}

\item \textit{Stage~1 (Exploration):} Following dataset preparation, 50 TPE trials are run for at most 150 epochs each with early stopping (patience of 25 epochs). Physics loss weights here are fixed to zero i.e, $\lambda_r = \lambda_s = \lambda_g = 0$, so that the search focuses entirely on the data loss. The model searches over the full broad hyperparameter space, including loss function selection over \{Huber, MAE, MSE, Log-Cosh\}, network width, depth, learning rate, weight decay, dropout, layer normalization, and batch size. Gradient clipping~\cite{Pascanu:2012} is applied throughout training using the constraint $\|\nabla_\theta \mathcal{L}\| \leq 1$, where $\nabla_\theta \mathcal{L}$ denotes the gradient of the loss function with respect to the model parameters. This prevents excessively large parameter updates and stabilizes the optimization when using the AdamW optimizer. The top 20\% of trials, ranked by validation yield error, are selected and promoted to Stage~2. For each selected trial, the model parameters are initialized using a warm-start strategy,
\begin{equation}
    \theta_0^{(2)} \;\leftarrow\; \theta_\star^{(1)},
    \label{eq:warmstart12}
\end{equation}

where $\theta_\star^{(1)}$ denotes the best-performing parameter set of the Stage~1 parent trial, and $\theta_0^{(2)}$ is the initial parameter state of the Stage~2 model. The warm-start replaces random weight initialization with a parameter configuration that already captures the basic spectral structure of the training data, allowing Stage~2 to focus on refining the physics constraints rather than relearning the data-driven component from scratch.

\item \textit{Stage~2 (Refinement):} The top 20\% of Stage~1 candidates are warm-started into Stage 2 through Eq. ~\eqref{eq:warmstart12}. Each child inherits a parameter state $\theta^{(1)}_*$ that already captures the dominant spectral structure. Stage 2 will guide the model in learning physics features without relearning the data-driven components from scratch. The architecture hyperparameters, such as width, depth, layer normalization, batch size, data loss function, and Huber $\delta$, are frozen from Stage 1 parent values; only the learning rate, weight decay, and dropout are refined within a narrow band centered around the parent's configuration. The selected models are introduced to the physics-informed loss with a narrow search over $\lambda_r,\,\lambda_s,\,\lambda_g \in [0,\,0.02]$, and the composite physics loss is evaluated every 5 epochs. Training is extended to 500 epochs with an early stopping patience of 40, during which 15 TPE trials are performed to optimize the physics-informed hyperparameters. Therefore, Stage 2 serves a dual purpose: first, it exploits the inherited weights and constrained architecture via warm-start, and second, it explores the previously unseen domain of physics loss coefficients. The top 8 candidates by validation yield error \eqref{eqn:yield_error} are promoted to Stage 3 using,
\begin{equation}
    \theta_0^{(3)} \;\leftarrow\; \theta_\star^{(2)}.
    \label{eq:warmstart23}
\end{equation}

\item \textit{Stage~3 (Final Search):}  Stage~3 inherits the full hyperparameter frozen. The search range is widened to $\lambda_r,\,\lambda_s,\,\lambda_g \in [0,\,0.1]$, which is five times broader than Stage 2, allowing the optimizer to probe stronger physics regularization now that the model weights are already conditioned on both data and a limited range of physics-loss coefficients explored in Stage~2. Training is extended to 1200 epochs with early-stopping patience of 60, and 8 TPE trials are run. The top 4 candidates are transferred to the post-refinement stage via

\begin{equation}
    \theta_0^{(\mathrm{post}),\,i}
    \;\leftarrow\; \theta_{\star,\,i}^{(3)},
    \qquad i = 1,\dots,k.
    \label{eq:warmstart_post}
\end{equation}
Here, $k$ represents the top candidates from Stage~3, with $k = 4$ in this study.

\begin{figure*}[t]
    \centering
\includegraphics[width=0.32\textwidth]{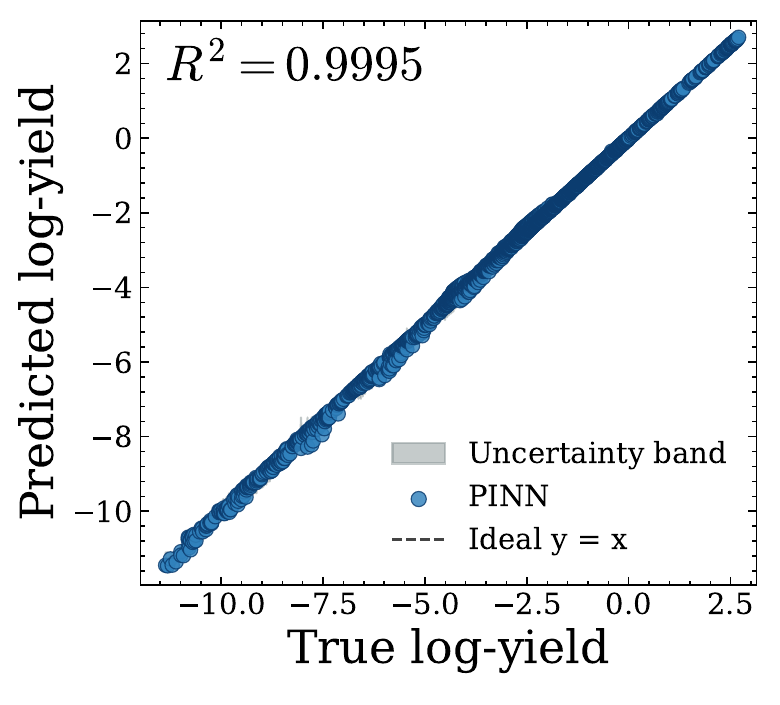}
  \includegraphics[width=0.32\textwidth]{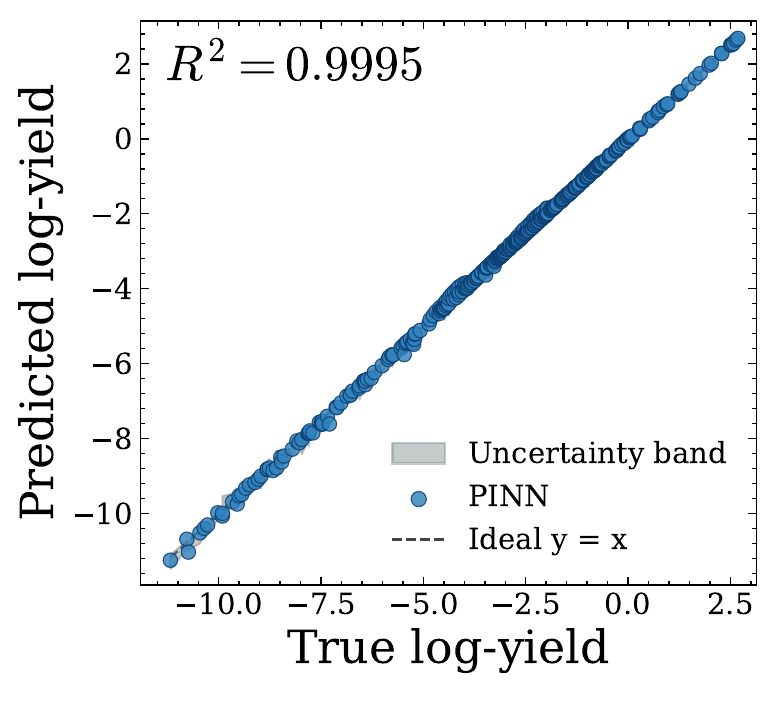}
  \includegraphics[width=0.32\textwidth]{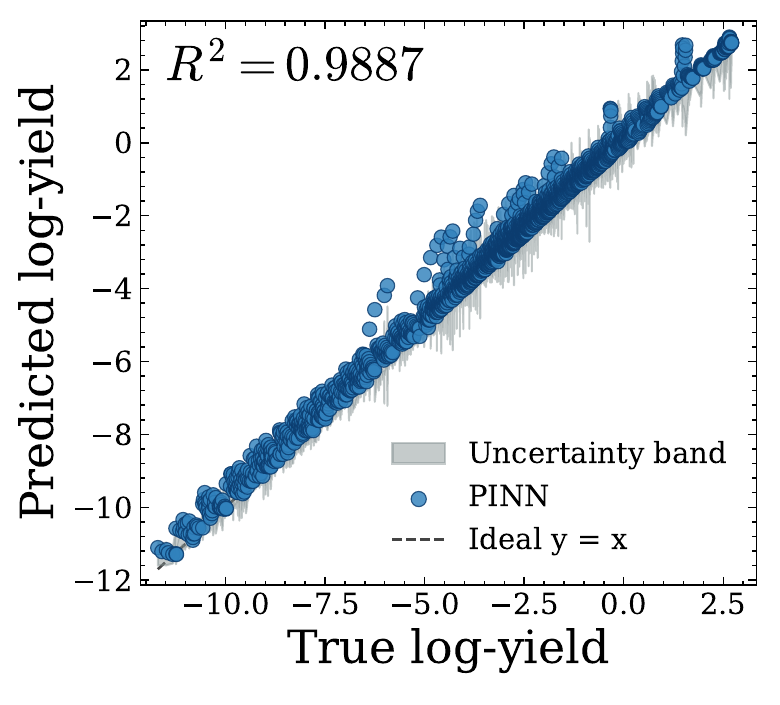}
    \captionof{figure}{Relation between the true and predicted output for the trained (left), interpolation (center), and extrapolation data (right).}
    \label{img:goodness_of_fit}
\end{figure*}
\end{itemize}
\textit{Post-Refinement:} In this stage, no further hyperparameter search is performed. All 4 models are trained to convergence for up to 4000 epochs with early stopping patience 500, it shares the same composite loss function as stage 3 with the hyperparameter inherited from its parent. The single model with the lowest validation yield error is selected and compared with XGBoost and LightGBM models.

\textit{Effect of staged warm-start on convergence:} 
The staged training procedure modifies the optimization trajectory by introducing the loss components sequentially. In Stage~1, the model is trained solely on data loss, allowing it to learn the dominant spectral structure. In Stage~2 and Stage~3, physics-informed loss terms are progressively introduced and strengthened, guiding the model toward physically consistent solutions. This staged conditioning reduces gradient competition between data and physics loss terms that would otherwise occur if all losses were applied simultaneously from random initialization. As a result, the optimization proceeds in a more stable and structured manner. In practice, for a fixed architecture and hyperparameter configuration, models trained with the staged warm-start approach achieve better performance than those trained from random initialization. This indicates that the optimization trajectory plays a critical role, as the staged procedure initializes the model parameters in a region of parameter space where physics-informed gradients can effectively refine the learned spectral structure.

\section{Results and Discussion}
\label{sec:results}

\subsection{Model Performance and Stability}

\begin{table*}[t]
\centering
\caption{Comparison of model performance. }
\label{tab:model_comparison}
\renewcommand{\arraystretch}{1.35}
\setlength{\tabcolsep}{4pt}
\begin{ruledtabular}
\begin{tabular}{llccccc}
 & \textbf{Model} & \textbf{Val MAE} & \textbf{Yield$_{\rm val}$ (\%)} & \textbf{Yield$_{\rm train}$ (\%)} & \textbf{Yield$_{\rm interp}$ (\%)} & \textbf{Yield$_{\rm extrap}$ (\%)} \\
\hline
\multirow{3}{*}{\rotatebox[origin=c]{90}{\footnotesize range}}
 & PINN
   & $0.01653^{+0.00858}_{-0.00319}$
   & $1.67^{+0.88}_{-0.32}$
   & $1.25^{+0.69}_{-0.59}$
   & $1.70^{+0.83}_{-0.43}$
   & $10.22^{+7.06}_{-4.39}$ \\
 & XGBoost
   & $0.02308^{+0.01704}_{-0.00481}$
   & $2.34^{+1.76}_{-0.49}$
   & $1.14^{+1.03}_{-0.61}$
   & $4.17^{+0.32}_{-0.12}$
   & $12.64^{+0.30}_{-0.13}$ \\
 & LightGBM
   & $0.02167^{+0.01282}_{-0.00456}$
   & $2.19^{+1.32}_{-0.47}$
   & $1.36^{+1.12}_{-0.63}$
   & $4.13^{+0.88}_{-0.13}$
   & $12.65^{+0.61}_{-0.18}$ \\
\hline
\multirow{3}{*}{\rotatebox[origin=c]{90}{\footnotesize best seed}}
 & PINN     & $0.01463$ & $1.47$ & $1.08$ & $1.78$ & $5.83$ \\
 & XGBoost  & $0.01826$ & $1.84$ & $0.97$ & $4.07$ & $12.57$ \\
 & LightGBM & $0.01712$ & $1.73$ & $0.94$ & $4.01$ & $12.50$ \\
\end{tabular}
\end{ruledtabular}
\end{table*}

The hyperparameter optimization for XGBoost and LightGBM was performed using standard Optuna TPE, while the PINN was optimized using the proposed staged hyperparameter search pipeline described in Sec .~\ref {sec:model_selection}. It was found that PINN performed better than both tree-based models, as evaluated by the yield error $ \left( \varepsilon_{\rm yield}  \right) $ in the trained, interpolation, and extrapolation regions. To assess sensitivity and verify the reproducibility of the model performance, each model was trained independently across 30 random seeds.

\begin{figure}
\includegraphics[width=0.48\textwidth]{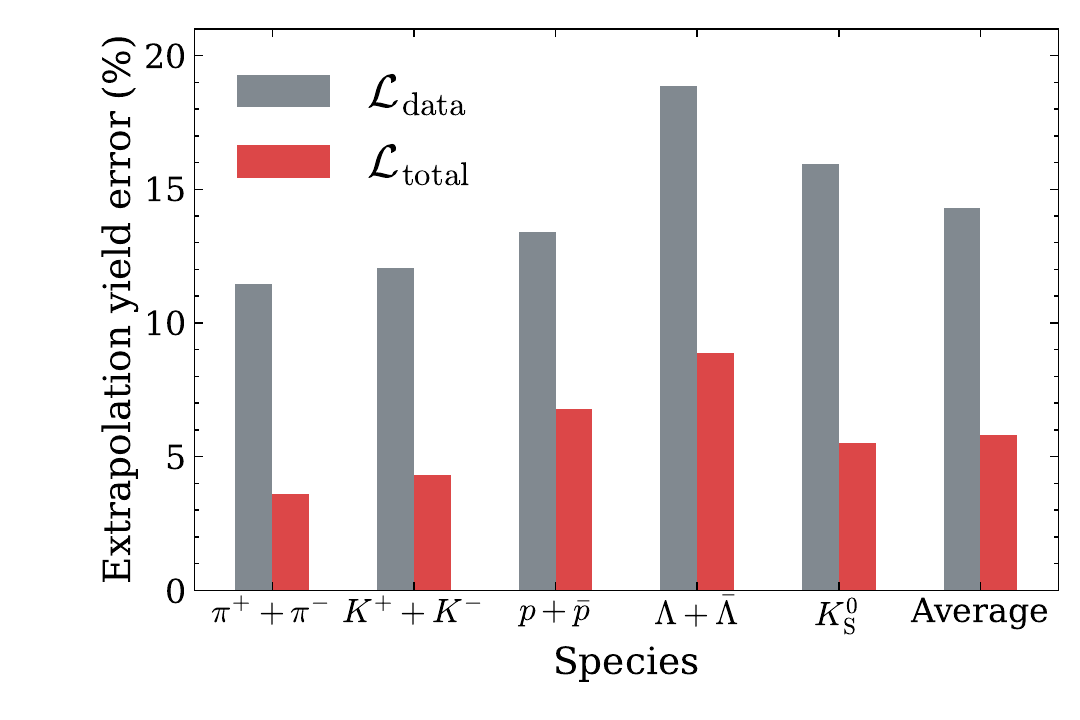}
    \caption{Comparison of extrapolation yield error across particle species for models trained using only $\mathcal{L_{\rm data}}$ and using $\mathcal{L_{\rm total}}$}
    \label{fig:ratiocompare}
\end{figure}

Table~\ref{tab:model_comparison} reports the mean and full observed range $(x_{\max}-x_{\min})$ of $\varepsilon_{\rm yield}$ across all 30 seeds, broken down by kinematic region. For PINN, the mean extrapolation error is $10.22^{+7.06}_{-4.39}\%$ across the ensemble. The observed spread reflects the sensitivity of the physics loss to the initialization of the weights set during model discovery at Stage 1, and some initializations converge to solutions that generalize well to unseen $\eta$ bins, while others converge to minimize the loss at the expense of extrapolation quality. Notably, the tree-based models achieve training yield error comparable to or lower than the PINN when compared against the best seed predictions, yet the prediction in the interpolation region, the loss ($\sim 4.1\%$), is a lot higher than the PINN and has even larger error in the case of extrapolation ($\sim 12.6\%$), which indicates overfitting to the training distribution. This behavior is expected, as the ensemble of tree-based models approximates the target function through a sum of piecewise constant base learners,  which sacrifices the continuous $\eta$-dependence of the spectra that the PINN captures through its differentiable functional form constrained by the physics loss terms. The extrapolation variance is $\lesssim 0.6\%$ because all learners converge to a stable minimum near a flat solution. Table~\ref{tab:model_comparison} reports the best performing seeds among all three models and their losses, among which seed 450 of PINN has shown the lowest interpolation $\varepsilon_{\rm yield}$ of 1.78\% and extrapolation $\varepsilon_{\rm yield}$ of about 5.83\%,  which is averaged over all the PIDs, $\eta$ and $N_{\rm ch}$ ranges. A brief comparison of all three models is shown in Fig.~\ref{fig:compare_ratio} of Appendix~\ref{app:1}, along with the architecture of the best-performing PINN model, which is shown in Table~\ref{tab:best_pinn}.

As used in Ref.~\cite{Shokr:2021ouh}, the coefficient of determination is given by,
\begin{equation}
    R^2 = \frac{\sum (\hat{y}_i - \bar{y})^2}{\sum (y_i - \bar{y})^2}
\end{equation}
where $y_i$ is the true value, $\bar{y}$ is the mean of $y_i$, and  $\hat{y}_i$ is the value predicted by the model. The relation between the truth and predicted values across all three cases is shown in Fig.~\ref{img:goodness_of_fit}. From the figure, it is clear that the PINN retains a reasonable agreement with the true values in all the kinematic regions, where the robustness of PINN slightly decreases when used for extrapolation, which is expected as the microscopic process changes towards higher pseudorapidity regions.

Fig.~\ref{fig:ratiocompare} shows the comparison between the yield error in the extrapolation region of the best model when it is trained completely only with the $\mathcal{L}_{\rm data}$ and when it is trained with $\mathcal{L_{\rm total}}$ as mentioned in Eq. ~\eqref{eq:loss_total}. The comparison shows that applying the physics-informed loss with the staged strategy yields lower yield error for all of the particles in the extrapolation region.

\begin{figure}[t]
    \centering    \includegraphics[width=0.45\textwidth]{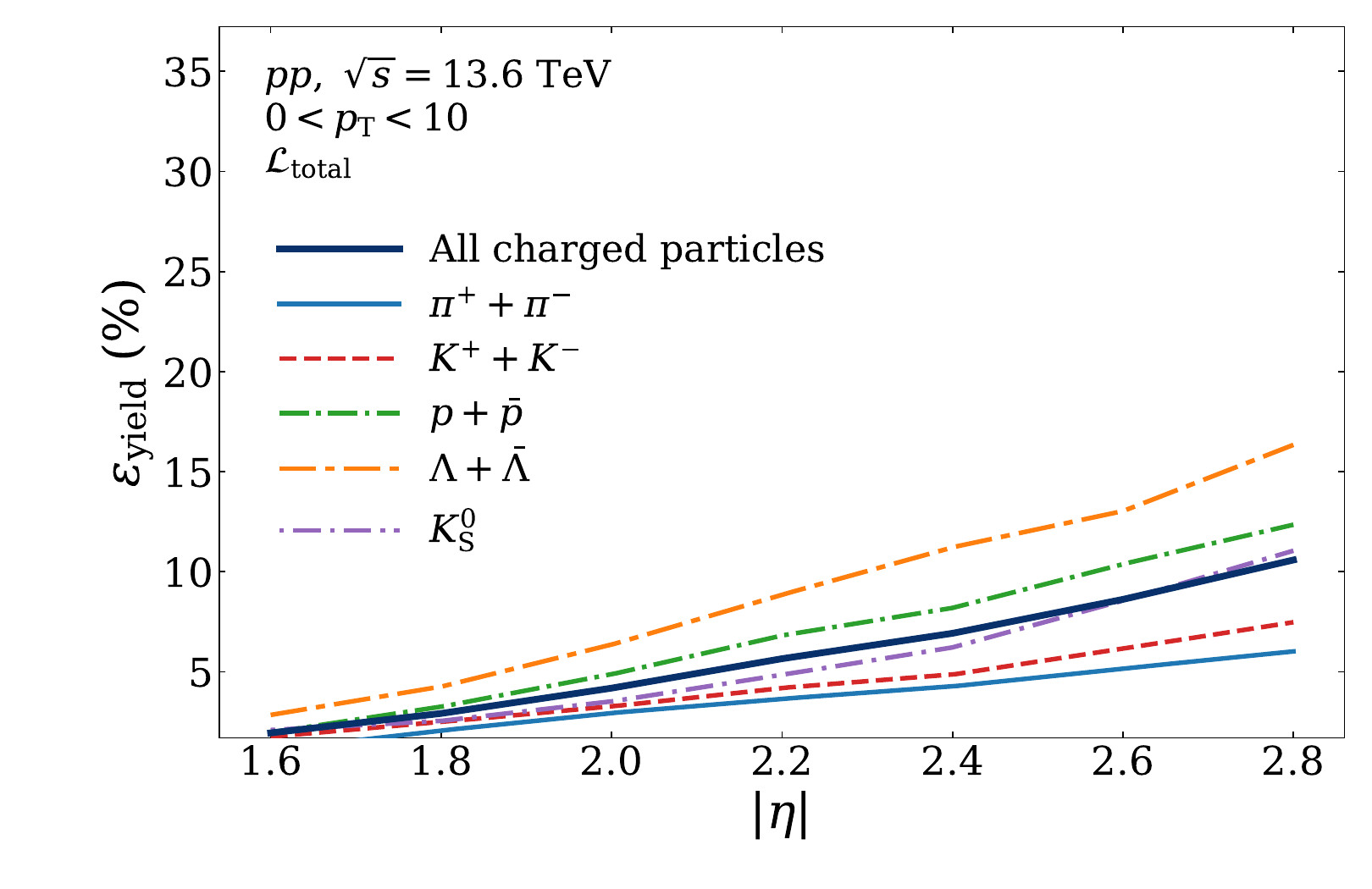}
    \includegraphics[width=0.45\textwidth]{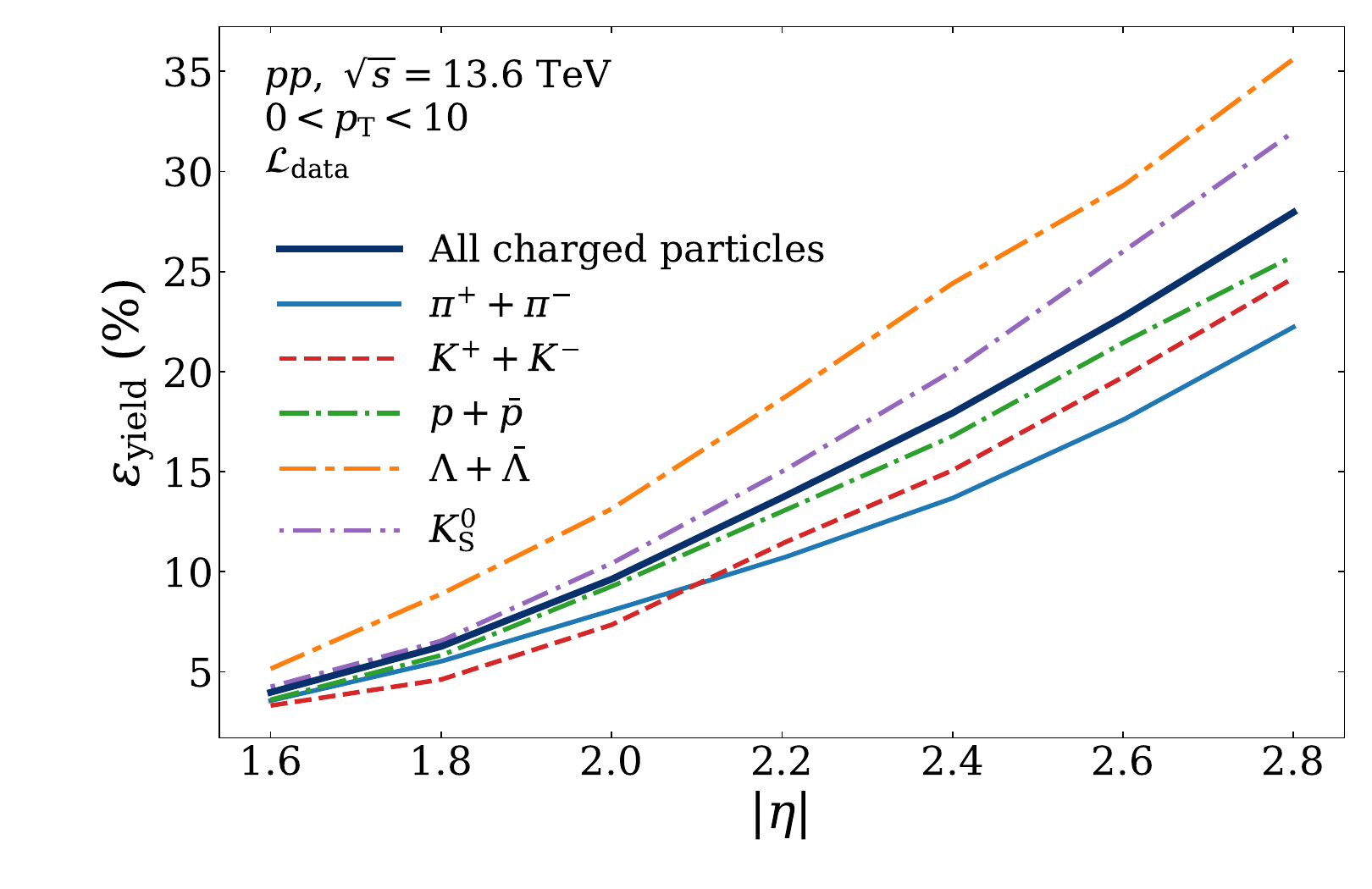}
    \caption{Extrapolation yield error as a function of $|\eta|$, averaged over all multiplicity classes.}
    \label{fig:2D_error}
\end{figure}

The error in the prediction from PINN for different particle species as a function of extrapolated $|\eta|$ region is shown in Fig.~\ref{fig:2D_error}, which is compared with that of all charged particles in pp collisions at $\sqrt{s}=13.6$ TeV. The comparison is made for $\mathcal{L}_{\rm data}$ only in the upper panel and $\mathcal{L}_{\rm total}$ in the lower panel of Fig.~\ref{fig:2D_error}. The comparison of upper and lower panels of Fig.~\ref{fig:2D_error} suggests that with the use of a physics-informed loss function, one can sufficiently reduce the error in the prediction. For $\Lambda+\bar{\Lambda}$, the error in  prediction reduces from $35\%$ using $\mathcal{L}_{\rm data}$ to $15\%$ with  $\mathcal{L}_{\rm total}$ at $|\eta|\approx 2.8$. While for the charged pions, the error in predictions reduces from $22\%$ to $6\%$ at a similar rapidity range. Additionally, the prediction uncertainties exhibit a dependence on particle mass, reflecting the imbalance in the training data: the model learns the features of abundant species such as pions more effectively than those of rarer particles like $\Lambda+\bar{\Lambda}$.

\begin{figure*}[p]
    \centering

    \begin{subfigure}{\textwidth}
        \centering
        \caption{Trained region}
        \includegraphics[width=0.32\textwidth]{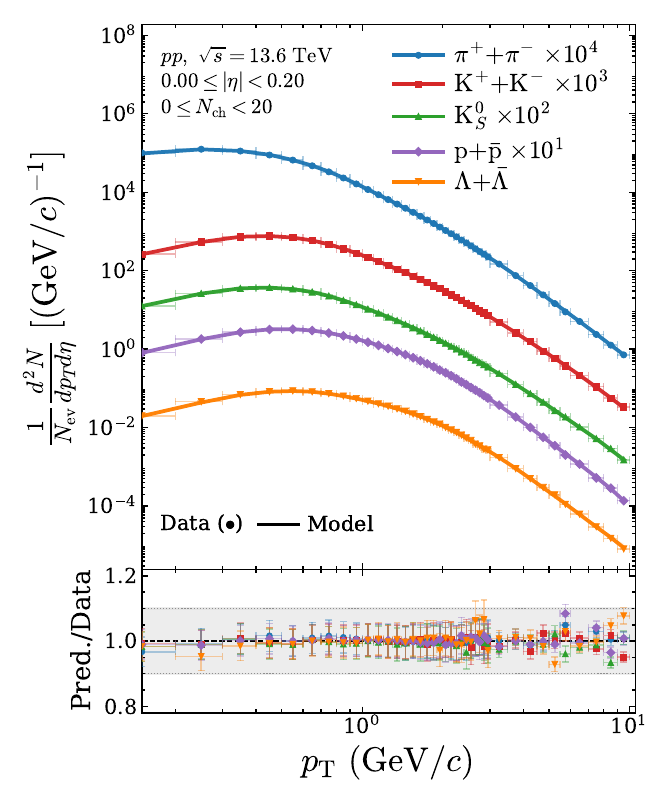}
        \includegraphics[width=0.32\textwidth]{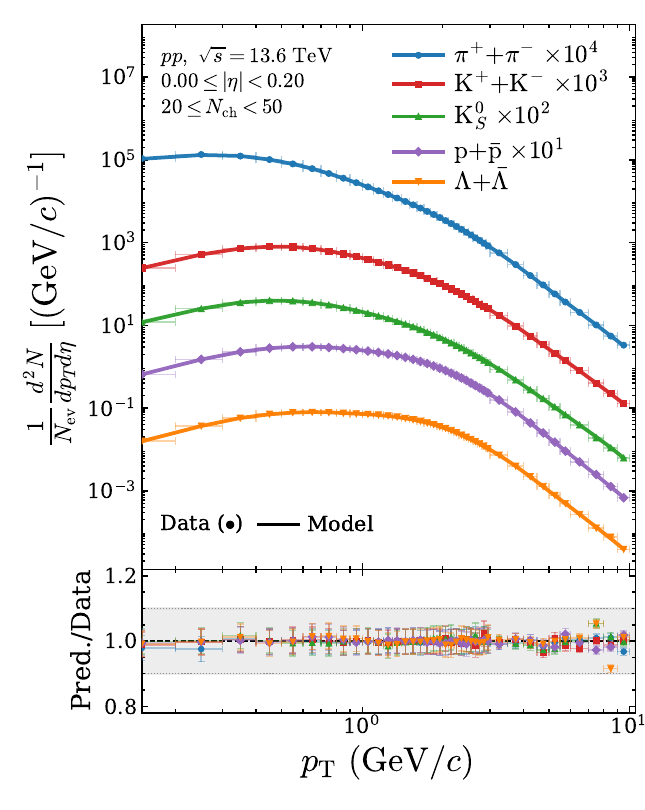}
        \includegraphics[width=0.32\textwidth]{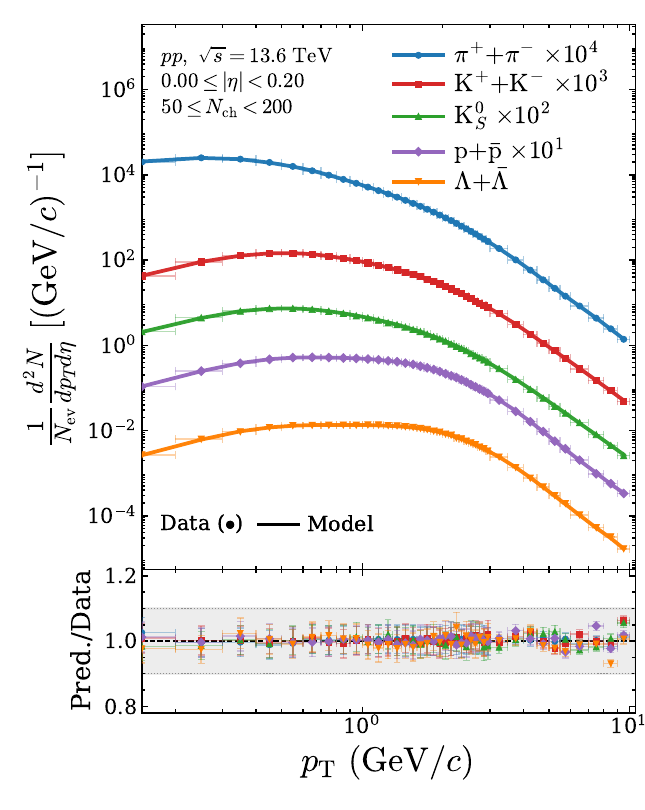}
        \label{fig:predict_train}
    \end{subfigure}
    \begin{subfigure}{\textwidth}
        \centering
        \caption{Interpolation region}
        \includegraphics[width=0.32\textwidth]{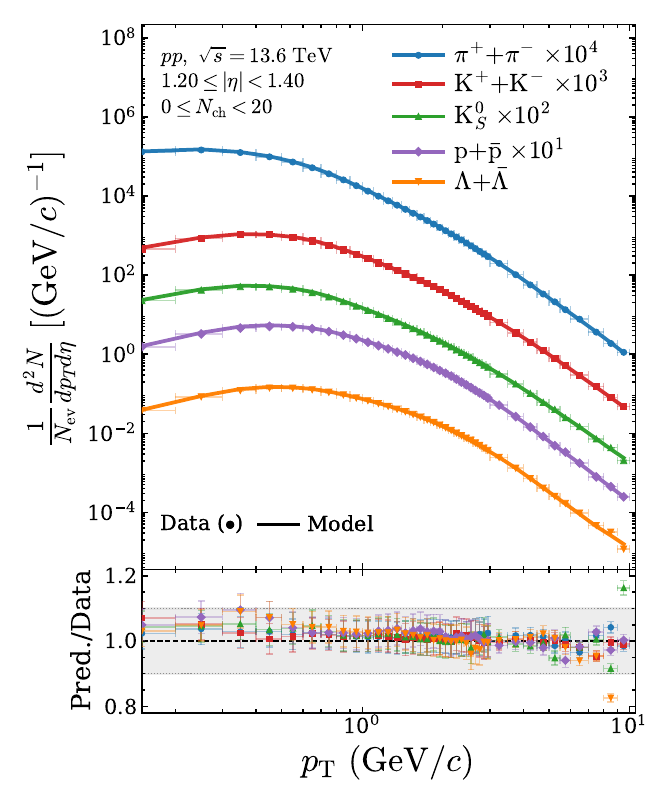}
        \includegraphics[width=0.32\textwidth]{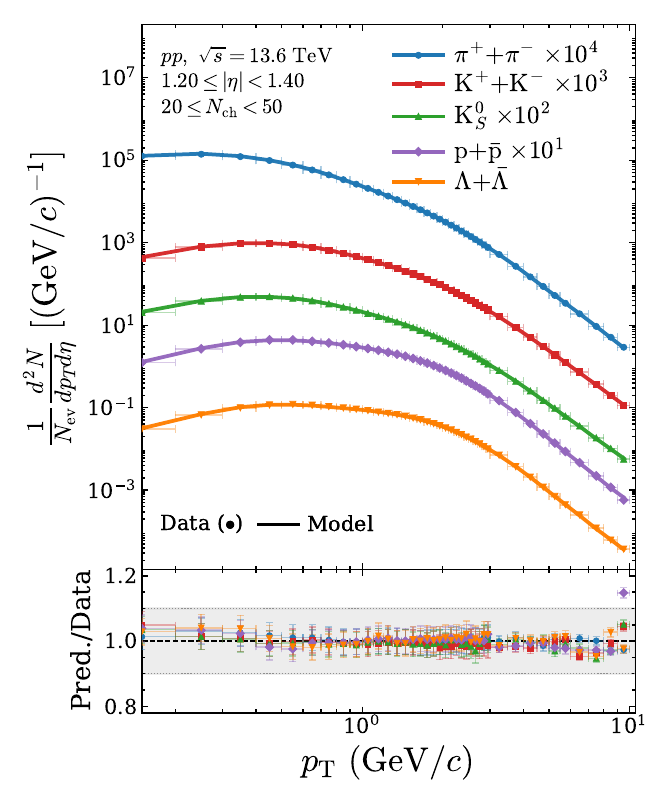}
        \includegraphics[width=0.32\textwidth]{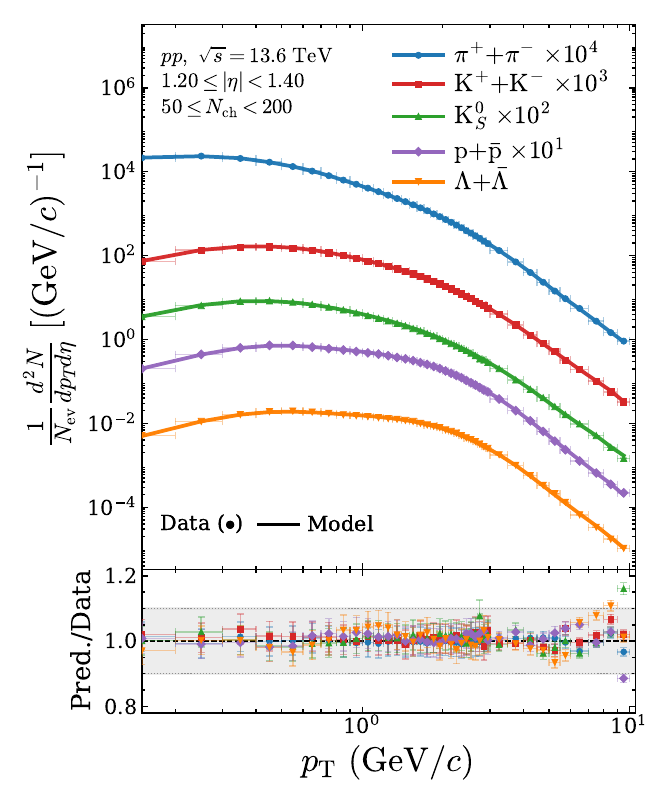}
        \label{fig:predict_interp}
    \end{subfigure}
    \begin{subfigure}{\textwidth}
        \centering
        \caption{Extrapolation region}
        \includegraphics[width=0.32\textwidth]{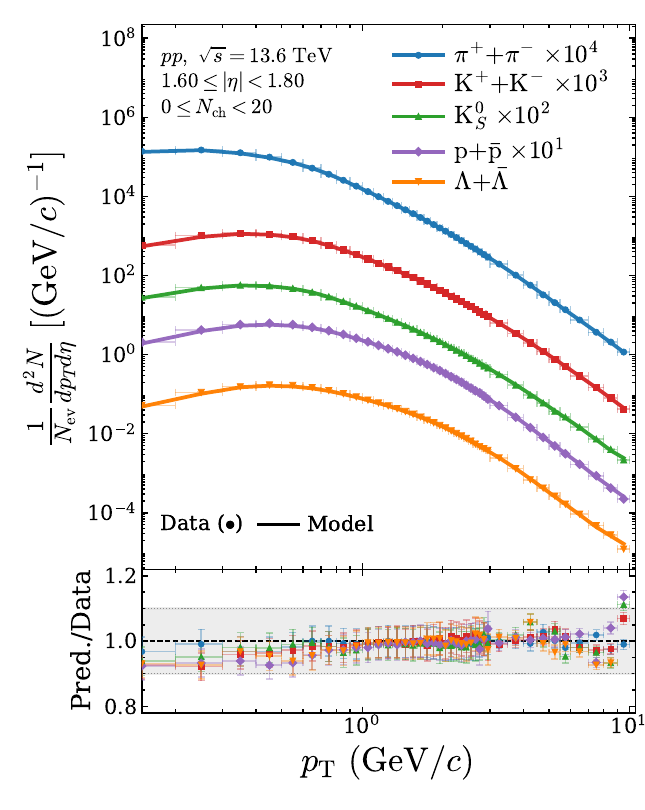}
        \includegraphics[width=0.32\textwidth]{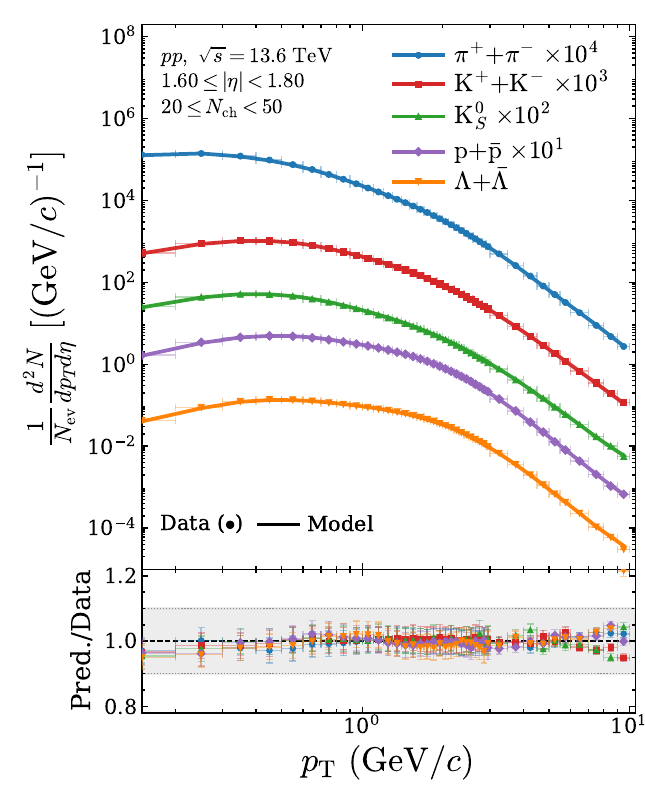}
        \includegraphics[width=0.32\textwidth]{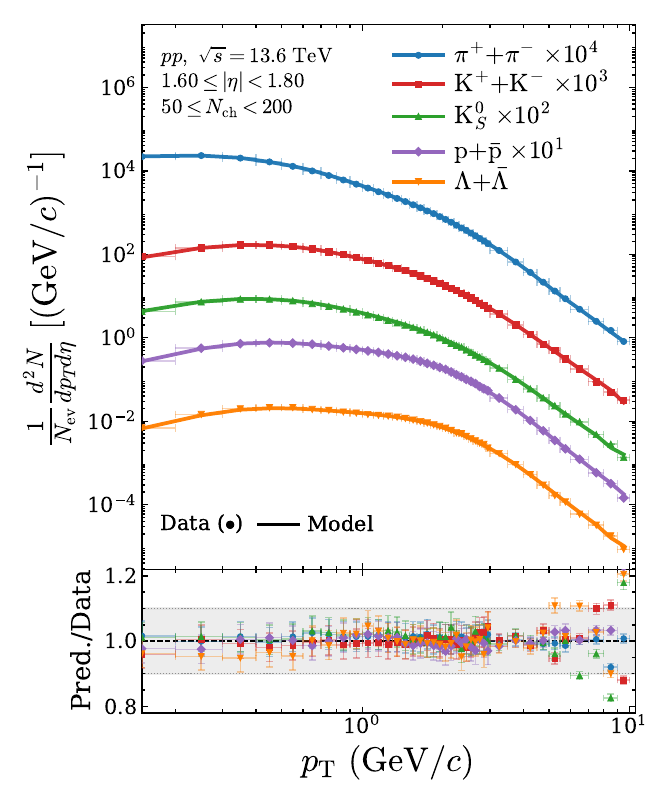}
        \label{fig:predict_extrap}
    \end{subfigure}

    \caption{PINN predictions across the trained, interpolation, and extrapolation regions for different multiplicity classes. The gray band in the ratio plot represents an error of $\pm10\%$. Here, `Data' refers to PYTHIA8, and `Model' refers to the best PINN model predictions.}
    \label{fig:pinn_predictions}
\end{figure*}

Fig.~\ref{fig:predict_train} shows the predicted transverse momenta spectra of the best performing PINN model across all three multiplicity classes $N_{\rm ch} \in (0,20)$, $(20,50)$, and $(50,200)$, from left to right, in the trained region, where the predicted spectra overlap with the true data. The deviation between the predicted and true values is very low, as shown in the ratio panels. In the interpolation region, as shown in Fig.~\ref{fig:predict_interp}, the model maintains this performance, which is expected since the interpolation gap spans only one $\eta$ bin. The gap could, in principle, be extended to further probe the generalization capability. However, in the present study, it is restricted to a single bin as an initial validity check of the model performance. Fig.~\ref{fig:predict_extrap} shows the model's prediction in the extrapolation region just away from the trained boundary across all the multiplicity classes. The prediction demonstrates that the model captures multiplicity dependence across $\eta$ bins, with minimal deviations observed for all $N_{\rm ch}$ classes, though the prediction error varies with the multiplicity class. One can infer the robustness of the trained PINN model using Fig.~\ref{fig:pinn_predictions}, which shows accurate predictions in different kinematic regions and multiplicity classes.

\subsection{Physics Validation}

\subsubsection{Particle ratios}

\begin{figure*}
    \centering

    \begin{subfigure}{\textwidth}
        \centering
        \caption{Interpolation region.}
        \label{fig:ratio_inter}
        \includegraphics[width=0.32\textwidth]{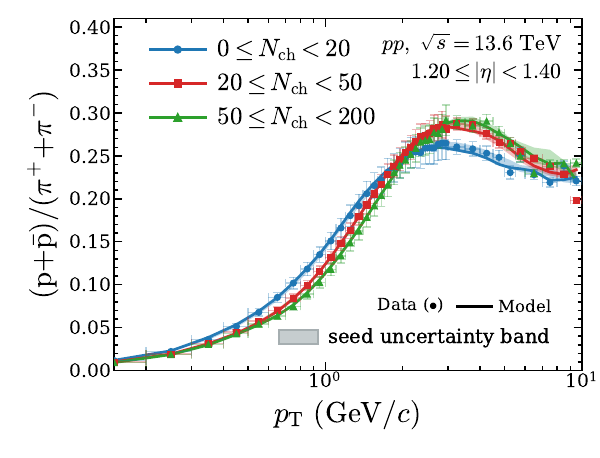}
        \includegraphics[width=0.32\textwidth]{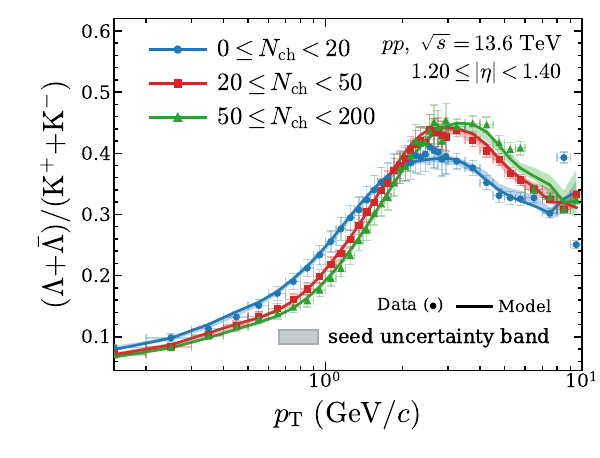}
        \includegraphics[width=0.32\textwidth]{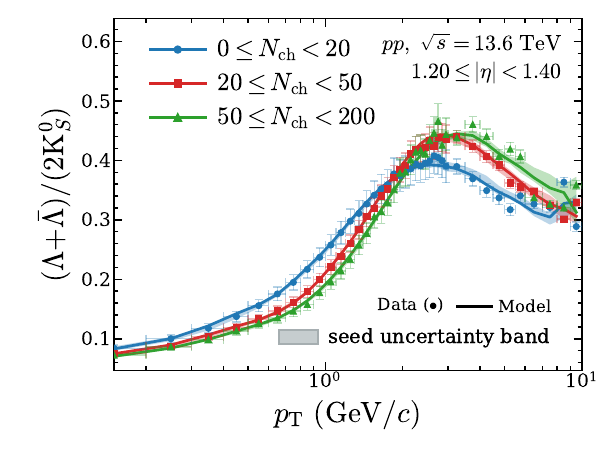}
    \end{subfigure}

    \begin{subfigure}{\textwidth}
        \centering
        \caption{Extrapolation region.}
        \label{fig:ratio_extrap}
        \includegraphics[width=0.32\textwidth]{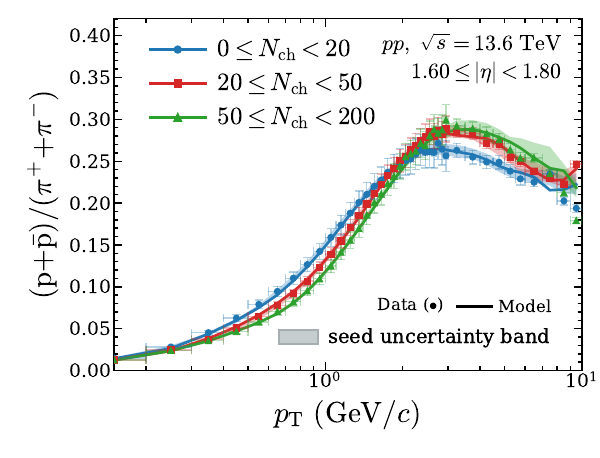}
        \includegraphics[width=0.32\textwidth]{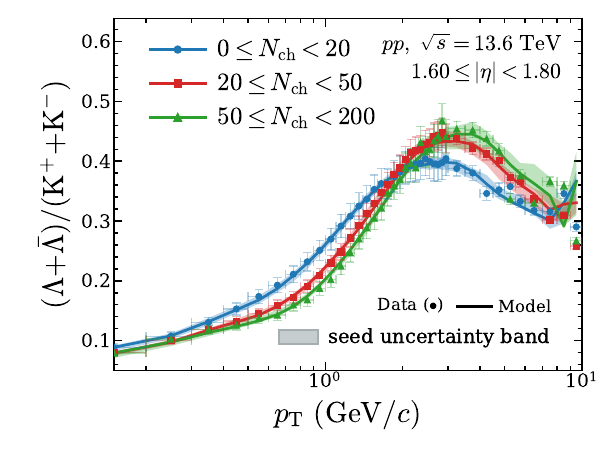}
        \includegraphics[width=0.32\textwidth]{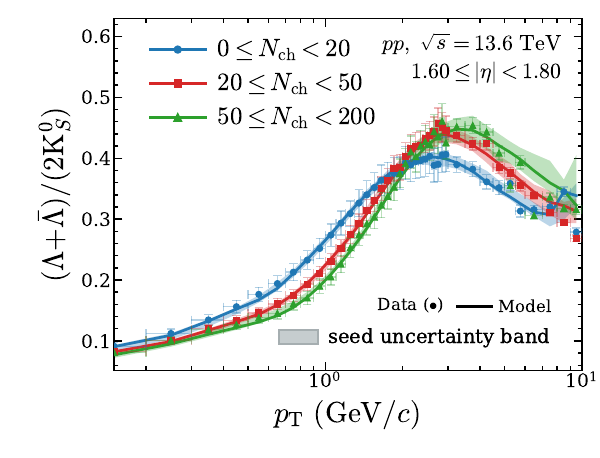}
    \end{subfigure}
    \caption{Comparison of particle ratio predictions with PYTHIA8 results in the interpolation and extrapolation regions across different multiplicity classes. Here, `Data' refers to PYTHIA8, and `Model' refers to the best PINN model predictions.}
    \label{fig:compare_ratio}
\end{figure*}

The collective expansion of the hot, dense medium formed in ultrarelativistic collision generates a radially outward flow of the particles that imparts a momentum boost to the produced hadrons~\cite{Siemens:1978pb, Schnedermann:1993ws}, known as radial flow.  Since the magnitude of this boost scales with the particle mass, heavier species acquire a larger momentum than lighter ones at the same flow velocity, resulting in the mass-dependent broadening of $\pT$ spectrum. The broadening is directly reflected in the $\pT$ dependent yield ratios of the particles of different masses in the presence of radial flow. The collective boost pushes heavier particles to higher $\pT$ more strongly than lighter ones. This results in the bump-like structure at the intermediate $\pT$ in the baryon to meson ratio. This bump shifts to higher $\pT$ as the flow velocity increases~\cite{MenonKavumpadikkalRadhakrishnan:2023cik,OrtizVelasquez:2013ofg}. The particle ratio, therefore, serves as an experimentally accessible probe of the mass ordering of the spectra and consequently of the radial flow velocity $\langle\beta_T\rangle$ of the system \cite{Retiere:2003kf, Schnedermann:1993ws}. 

As $N_{\rm ch}$ increases, the enhanced Multi-Parton Interaction (MPI) activity generates a larger string density~\cite{ALICE:2016fzo}. Through Color Reconnection (CR), this induces a stronger collective-like boost, leading to the characteristic mass-ordering structure in baryon-to-meson ratios, whose magnitude depends on the hadron mass~\cite{MenonKavumpadikkalRadhakrishnan:2023cik,OrtizVelasquez:2013ofg}. 

In the present study, the ratio consistency loss $\mathcal{L}_{\rm ratio}$ explicitly enforces $(K^++K^-)/(\pi^++\pi^-)$, $(p+\bar{p})/(\pi^++\pi^-)$, and $(\Lambda+\bar{\Lambda})/K^0_S$ during 
training. Fig~\ref{fig:ratio_inter} shows the predicted and truth particle ratios as a function of $\pT$ in the interpolation region  $1.20\leq |\eta|<1.40$ across all three multiplicity classes. The error band in the figure corresponds to the different seed PINN model predictions. The model reproduces the multiplicity dependence of all three ratios, capturing increasing baryon to meson ratios with $N_{\rm ch}$ and the correct $\pT$ dependent shape of each ratio. In the extrapolation region as shown in \ref{fig:ratio_extrap}, the model continues to reproduce the correct multiplicity ordering of the ratios, confirming that the spectral prediction quality is maintained beyond the training boundary across the multiplicity classes. Since these ratios encode the mass ordering of the $\pT$ spectra and probe the collective flow-like dynamics of the system ~\cite{Retiere:2003kf, Schnedermann:1993ws, OrtizVelasquez:2013ofg}, their accurate reproduction serves as a physically motivated validation of the model's prediction capability. 
\begin{figure*}
     \begin{subfigure}{\textwidth}
        \centering
        \includegraphics[width=0.32\textwidth]{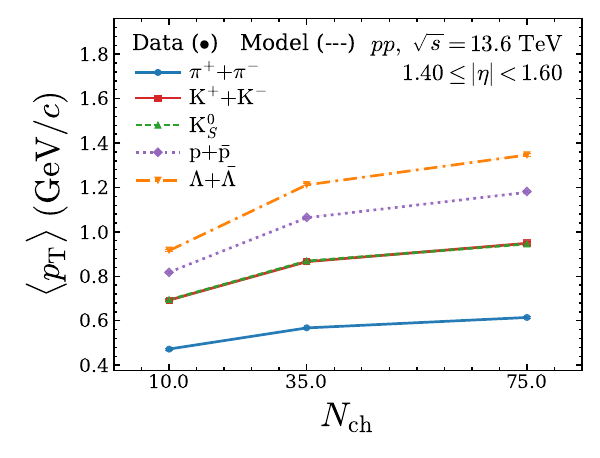}
        \includegraphics[width=0.32\textwidth]{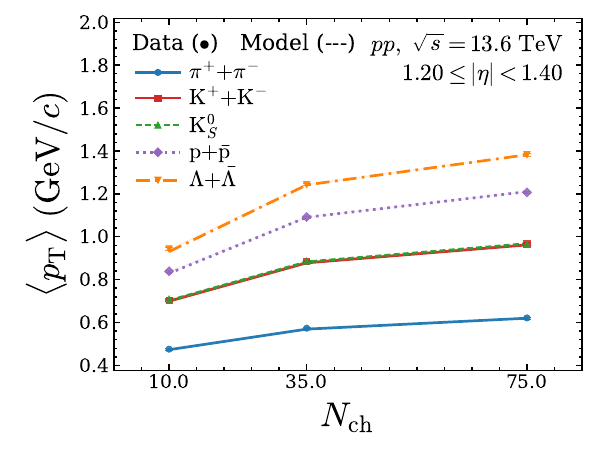}
        \includegraphics[width=0.32\textwidth]{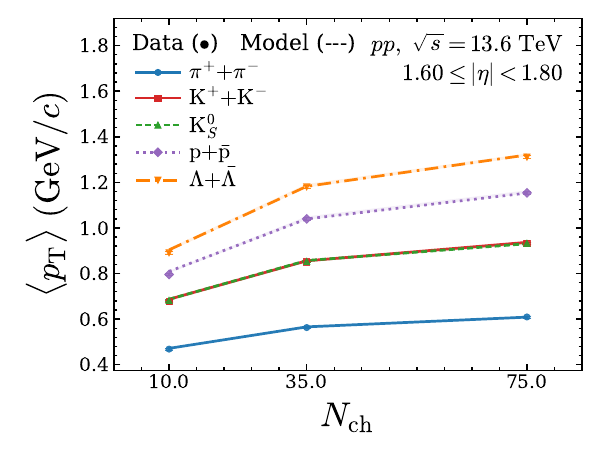}
    \end{subfigure}
    \caption{Mean transverse momentum as a function of charged-particle multiplicity $N_{\rm ch}$ at midrapidity. The PYTHIA8 data (markers) and PINN predictions (lines) for all five identified species in the trained (left), interpolation (center), and extrapolation (right) $\eta$ ranges. Here, `Data' refers to PYTHIA8, and `Model' refers to the best PINN model predictions.
    }
    \label{fig:mean_pt}
\end{figure*}

\subsubsection{Mean transverse momentum}
The mean transverse momentum $\left( \langle \pT\rangle \right)$ as a function of charged-particle multiplicity ($N_{\rm ch}$) provides complementary and integrated validation of the model's spectral prediction quality. Driven by the growing MPI activity and color reconnection, the $\langle \pT\rangle$ is expected to increase with $\langle N_{\rm ch}\rangle$ in pp collisions~\cite{ALICE:2013rdo, ALICE:2020nkc, ALICE:2018pal}. This rise is more pronounced for heavier species the mass dependent boost from color reconnection causes $\langle \pT \rangle$  to increase faster for protons than kaons and pions~\cite{OrtizVelasquez:2013ofg, MenonKavumpadikkalRadhakrishnan:2023cik}, this produces a characteristic mass ordering $\langle \pT \rangle_\Lambda > \langle \pT \rangle_p > \langle \pT \rangle_K > \langle \pT \rangle_\pi$ at all multiplicity classes. Fig.~\ref{fig:mean_pt} shows $\langle \pT \rangle$ as a function of $N_{\rm ch}$ for all five hadron species across the trained, interpolation, and extrapolation $\eta$ ranges for the truth and predicted spectra. The model correctly reproduces the rising trend without being explicitly trained on the CR mechanism of PYTHIA8 and is consistent with the radial-flow-like behavior observed in PYTHIA8 data.

\subsubsection{Kinetic freeze out parameters validation}
In ultrarelativistic collisions, the spacetime evolution of the system proceeds through several distinct phases. Starting from the pre-equilibrium phase, the system evolves into the QGP phase, followed by the mixed phase, chemical freeze-out, and the hadron gas phase ~\cite{Braun-Munzinger:1994ewq, Huovinen:2001cy}. Finally, when the interaction rate drops sufficiently, the momenta of the produced hadrons cease to change, reaching the kinetic freeze-out stage. The radial flow, built up during the QGP and mixed phases, is carried through the hadronic phase and imprints itself on the final state momentum distributions at kinetic freeze-out ~\cite{Siemens:1978pb, Schnedermann:1993ws}, providing a collective boost to the surviving hadrons. This effect is reflected in the transverse momentum spectra of the produced hadrons as an enhancement in the collective expansion velocity of the hadronic system~ \cite{MenonKavumpadikkalRadhakrishnan:2023cik, ALICE:2022wpn}. To assess the sensitivity of the best performing model to such physical quantities encoded in the shape of the $\pT$ spectra, we extract the kinetic freeze out temperature $T_{\rm kin}$ and the mean radial flow velocity $\langle\beta_T\rangle$ by performing a simultaneous Boltzmann-Gibbs blast-wave (BGBW) fit to the $(\pi^++\pi^-)$, $(K^++K^-)$, and $(p+\bar{p})$ spectra over the $\pT$ ranges $[0.5, 1.2]$~GeV/$c$ for pions, $[0.3, 1.5]$~GeV/$c$ for kaons, and $[0.8, 2.0]$~GeV/$c$ for protons which is the same as used in Ref~\cite{MenonKavumpadikkalRadhakrishnan:2023cik, Ortiz:2015ttf}. Boltzmann-Gibbs blast-wave function applied in the midrapidity is given by \cite{Schnedermann:1993ws, BRAHMS:2016klg}:
\begin{align}
\left.\frac{d^2 N}{d\pT dy}\right|_{y=0}
&= C\, \pT m_T \int_0^{R_0}
\Biggl[
r D(r)\, dr \times \nonumber \\
&\qquad
K_1\!\left(\frac{m_T \cosh \rho}{T_{\rm kin}}\right)
I_0\!\left(\frac{\pT \sinh \rho}{T_{\rm kin}}\right)
\Biggr]
\end{align}

\noindent
where the modified Bessel functions are given by
\begin{align}
K_1\!\left(\frac{m_T \cosh \rho}{T_{\text{kin}}}\right)
&= \int_{0}^{\infty} \cosh y \,
\exp\!\left(-\frac{m_T \cosh y \cosh \rho}{T_{\text{kin}}}\right) dy,
\label{eq:K1}
\\
I_0\!\left(\frac{\pT \sinh \rho}{T_{\text{kin}}}\right)
&= \frac{1}{2\pi} \int_{0}^{2\pi}
\exp\!\left(\frac{\pT \sinh \rho \cos \phi}{T_{\text{kin}}}\right) d\phi.
\label{eqn:bgbw}
\end{align}

here, $D(r)$ is the nuclear density profile. In this study, a hard sphere 
profile is assumed taking from Ref~\cite{MenonKavumpadikkalRadhakrishnan:2023cik}, defined as

\begin{equation}
    D(r) = 
    \begin{cases}
        1, & r \leq R_0 \\
        0, & r > R_0
    \end{cases}
\end{equation}

where $r$ is the radial distance and $R_0$ is the maximum hard sphere radius of the source at freeze out. $\rho=\tanh^{-1}\beta_T$ where $\beta_T=\beta_s \xi^n $~\cite{Schnedermann:1993ws, Huovinen:2001cy, Braun-Munzinger:1994ewq, Tang:2011xq} and $\beta_T$ is the radial flow velocity, $\beta_s$ is the maximum surface velocity, $\xi=(r/R_0)$, where $n$ is a free parameter which describes the expansion profile. $\langle \beta_T \rangle$ gives the mean transverse velocity, which gives the information of the boost on the hadrons expanding outwards and is defined as~\cite{PHENIX:2003wtu}:
\begin{equation}
\langle \beta_T \rangle = \frac{\int \beta_s \xi^n \xi \, d\xi}{\int \xi \, d\xi} = \left(\frac{2}{2+n}\right) \beta_s
\end{equation}
The modified Bessel's function $K_1(z)$ in Eq.~\ref{eqn:bgbw} results from the integration in the range $-\infty$ to $+\infty$ over $\eta$ assuming boost invariance. In the forward rapidity, this assumption of boost is not valid, hence $K_1(z)$ is replaced by the integration over a finite range $\eta$ given by $g(z)$ as \cite{BRAHMS:2016klg}:
\begin{equation}
g(z) = \int_{\eta_{\min}}^{\eta_{\max}} \cosh(\eta - y)\, e^{-z \cosh(\eta - y)}\, d\eta
\
\label{eqn:gz}
\end{equation}
here the value of $y$ lies exactly at the center of polar angles corresponding to $\eta$ bin range given by $\eta_{min}$ and $\eta_{max}$. Substituting Eq. ~\ref{eqn:gz} in Eq.~\ref{eqn:bgbw} gives:
\begin{align}
\left.\frac{d^2 N}{d\pT dy}\right|_{y=0}
&= C\, \pT m_T \int_0^{R_0}
\Biggl[
r\, dr \times \nonumber \\
&\qquad
g\!\left(\frac{m_T \cosh \rho}{T_{\rm kin}}\right)
I_0\!\left(\frac{\pT \sinh \rho}{T_{\rm kin}}\right)
\Biggr]
\label{eqn:bgbw}
\end{align}

\begin{figure}
    \centering
    \includegraphics[width=0.5\textwidth]{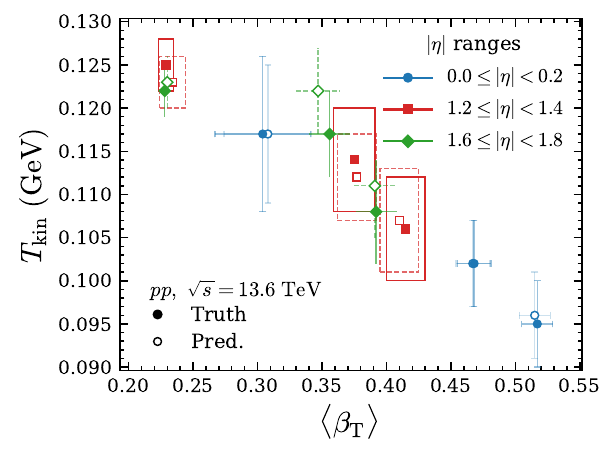}
    \caption{Kinetic freeze out temperature versus mean transverse radial flow velocity extracted from simultaneous BGBW fits to the truth and predicted spectra for multiplicity classes $N_{\rm ch} \in (0,20)$, $N_{\rm ch} \in (20,50)$, and $N_{\rm ch} \in (50,200)$ across trained, interpolation, and extrapolation $\eta$ ranges. Here, `Truth' refers to PYTHIA8, and `Pred' refers to the best PINN model predictions.}
    \label{fig:bparameters}
\end{figure}

The BGBW function is applied independently to both the truth and predicted spectra using the global $\chi^2$ minimisation performed simultaneously across the three species, $(\pi^++\pi^-)$, $(K^++K^-)$, and $(p+\bar{p})$, following the same method as described in Ref.~\cite{ Schnedermann:1993ws,Prasad:2021bdq,ALICE:2013mez,MenonKavumpadikkalRadhakrishnan:2023cik}, and the extracted freeze out parameters are compared directly. Fig.~\ref{fig:bparameters} shows the correlation between the extracted $T_{\rm kin}$ and $\langle\beta_T\rangle$ for the truth and predicted spectra across all three multiplicity classes and $\eta$ bins. In the trained region, the predicted spectra reproduce the truth spectral shapes sufficiently well that the independently extracted $\langle\beta_T\rangle$
 and $T_{\rm kin}$ values agree closely, with the predicted truth symbols overlapping within the uncertainty range across all the multiplicity classes. In the interpolation region, the model predicts both the parameters without significant deviation, following the same anticorrelation trend between both of the variables, which agrees with Refs.~\cite{Schnedermann:1993ws, ALICE:2013mez, MenonKavumpadikkalRadhakrishnan:2023cik}. 

 In the extrapolation region, $\langle\beta_T\rangle$ is well reproduced across all the multiplicity classes within the uncertainty ranges. The ratio loss $\mathcal{L}_{\mathrm{ratio}}$ explicitly enforces the ratios across different species of different mass, since variation in the temperature affects lighter species more strongly in low $\pT$, while the collective flow boost shifts the spectra of heavier particles to higher $\pT$ \cite{Retiere:2003kf, Schnedermann:1993ws}, the ratio loss enforces the mass ordering across species, which relates to the information on $\langle\beta_T\rangle$, resulting in the consistent reproduction across all multiplicity classes well within uncertainty range. $T_{\rm kin}$ is sensitive to the slope of the individual spectra in the low $\pT$ regime, more specifically, the per species logarithmic slope in the transverse mass $m_T = \sqrt{\pT^2 + m_s^2}$ evaluated at low $\pT$~\cite{Retiere:2003kf, Schnedermann:1993ws}.

 While the spectral smoothness loss $\mathcal{L_{\rm smooth}}$ and the shape loss $\mathcal{L}_{\rm shape}$ provide indirect constraints on the spectral curvature and gradient profile across the full $\pT$ range, these losses operate in $\pT$ space over the entire kinematic range and are not restricted to the low $\pT$ regime which determines the $T_{\rm kin}$ values. The data loss $\mathcal{L}_{\rm data}$ plays the primary role in enforcing the per species spectral points along the predicted trend, however, the residual deviation in the predicted spectra in the extrapolation region translates into a larger deviation in the extracted $T_{\rm kin}$ for the $N_{\rm ch} \in (20, 50)$ multiplicity class, as can be seen in Fig.~\ref{fig:bparameters}. The extracted values of $\langle\beta_T\rangle$  and $T_{\rm kin}$ for all $\eta$ ranges up to $1.8 \leq |\eta| < 2.0$ are listed in Table~\ref{tab:blastwave_all_eta}, as beyond this range the yield error increases linearly with $|\eta|$ bin, reducing the reliability of the extracted freeze out parameters. The extracted values nevertheless remain within the uncertainty range across all multiplicity classes and $\eta$ ranges, confirming that the model is successfully able to predict the identified-particle $\pT$ spectra across all multiplicity classes and $\eta$ ranges, and the reproduction of the 
kinetic freeze-out parameters $\langle\beta_T\rangle$ and 
$T_{\rm kin}$ serves as a physical validation of the model's predictive capability.

 \section{Summary and Outlook}
\label{sec:summary}
In this work, as a methodological proof of concept, we introduce a staged hyperparameter optimization strategy tailored for physics-informed neural networks to reconstruct hadron yields in unmeasured pseudorapidity regions. The model is trained on PYTHIA8-generated $pp$ collisions at $\sqrt{s}=13.6$ TeV and is designed to learn the dependence of particle production on kinematic and event-level variables, including $\pT$, $\eta$, and charged-particle multiplicity. Physically consistent predictions are ensured by employing PINNs with domain-specific loss functions that account for spectral smoothness, shape consistency, and particle yield ratios. These constraints are introduced through a staged hyperparameter optimization strategy, where the model is first trained using data loss alone and subsequently refined under progressively stronger physics-informed regularization. This staged warm-start procedure stabilizes the optimization and reduces gradient competition between data-driven and physics-based objectives. 

The model successfully reproduces the $\pT$ spectra, particle ratios, and $ \langle \pT \rangle$ as a function of multiplicity. In addition, we extracted kinetic freeze-out parameters $T_{\rm kin}$ and $\langle \beta_T \rangle$ from predicted spectra using Boltzmann-Gibbs blast-wave fits, and compared them to the values extracted from the PYTHIA8-generated spectra using the same procedure. The extracted parameters are in good agreement within uncertainty, across training, interpolation, and extrapolation regions. This establishes that the model learns the physical structure of the spectra and not merely their point-wise values. A comparison in which the physics-informed loss terms are removed shows that the full composite loss yields better extrapolation performance than training on data loss alone.

Beyond the present study, the framework extends naturally in multiple directions. First, the deviation observed in the extracted $T_{\rm kin}$ at intermediate multiplicities indicates that the loss function needs to be modified to target the low-$\pT$ slope and smoothness, motivating the addition of $\pT$- and multiplicity-dependent loss terms to better capture the underlying mechanism. Second, the methodology can be applied to reconstruct exotic particles, multi-strange baryons, and resonances, where one naturally encounters limited statistics. Here, the physics-informed loss structure can be extended to incorporate additional conservation laws and daughter-decay constraints. Third, the staged HPO protocol generalizes beyond spectrum reconstruction and can serve as a baseline methodology for training ML models under composite physics-motivated losses. Lastly, this framework can be further developed and trained on smeared detector-like data, which, in principle, should enable its use with realistic experimental data. This work establishes physics-informed neural networks, together with the staged hyperparameter optimization strategy, as a robust and reliable framework for reconstructing particle spectra in unmeasured regions.

\section*{Acknowledgment}
K.G. acknowledges financial support from the Prime Minister's Research Fellowship (PMRF), Government of India. S.P. acknowledges the support from the Hungarian National Research, Development and Innovation Office (NKFIH) under the contract numbers NKFIH NKKP ADVANCED\_25-153456, 2025-1.1.5-NEMZ\_KI-2025-00005, 2024-1.2.5-TET-2024-00022, and the usage of Wigner Scientific Computing Laboratory (WSCLAB). The authors (K.G., S.P., and R.S.) acknowledge funding from the DAE-DST, Government of India, under the mega-science project “Indian participation in the ALICE experiment at CERN,” bearing Project No. SR/MF/PS-02/2021-IITI(E-37123). This work has been carried out as a part of the M.Sc. (Physics) thesis by Rishabh Gupta at IIT Indore.

\appendix
\onecolumngrid
\section{Comparison of models}
\label{app:1}
In Fig.~\ref{fig:compare_ratio}, we compare the predicted $\pT$ spectra with the PYTHIA8-generated data across three pseudorapidity regions: the training region (left), an interpolation region (middle), and an extrapolation region (right), along with the corresponding model-to-data ratios. In the training region, all three models (PINN, XGBoost, and LightGBM) reproduce the spectra well, with ratios consistent with unity across the full $\pT$ range. In the interpolation region, while the PINN remains close to unity, XGBoost and LightGBM exhibit systematic discrepancies at low $\pT$, reaching up to 25\%. A similar trend is observed in the extrapolation region, where the PINN continues to perform reasonably well, whereas the tree-based models show deviations at low $\pT$ and increased fluctuations at higher $\pT$. Overall, the results indicate that while all models perform comparably in the training domain, the PINN demonstrates better stability and generalization, particularly in interpolation and extrapolation regions.
\begin{figure*}[h]
    \centering

    \includegraphics[width=0.32\textwidth]{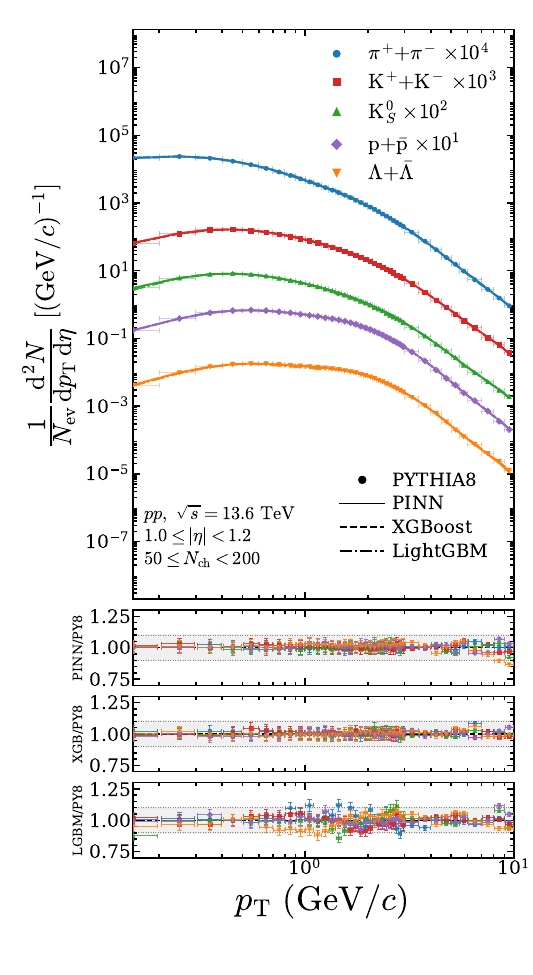}
    \hfill
    \includegraphics[width=0.32\textwidth]{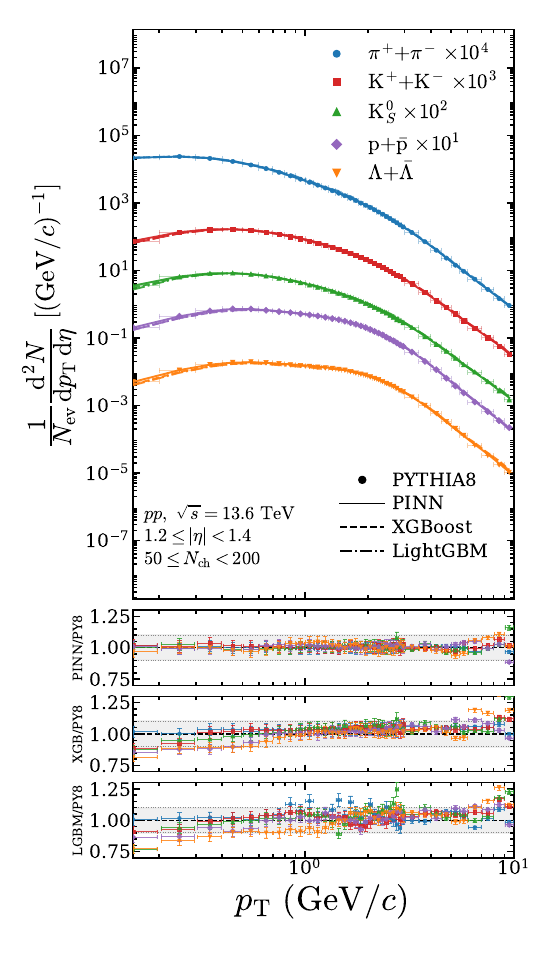}
    \hfill
    \includegraphics[width=0.32\textwidth]{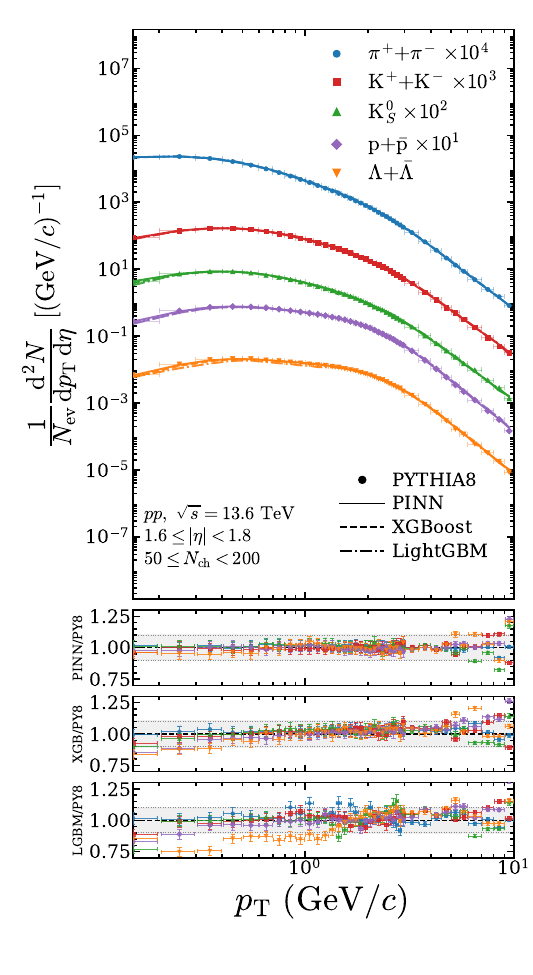}
    \caption{Comparison of $\pT$-spectra predictions from PINN, XGB, and LGBM in the trained (left), interpolation (middle), and extrapolation (right) pseudorapidity regions in pp collisions at $\sqrt{s}=13.6$ TeV.}
    \label{fig:compare_ratio}
\end{figure*}
\begin{table}
\centering
\caption{Best PINN configuration for the minimum-extrapolation seed used in the main model comparison lower block.}
\label{tab:best_pinn}
\begin{tabular}{l l}
\toprule
\textbf{Parameter} & \textbf{Value} \\
\midrule
\multicolumn{2}{l}{\textit{Architecture}} \\
Model type & Fully connected neural network \\
Input features & $|\eta|$, $N_{\rm ch}$, $\pT$, PID (one-hot encoded) \\
Input dimension & 8 \\
Hidden layers & 3 \\
Hidden units per layer & 134 \\
Activation function & $\tanh$ \\
Layer normalization & Enabled \\
Output & $\ln(\mathrm{yield})$ \\
\midrule
\multicolumn{2}{l}{\textit{Training Hyperparameters}} \\
Optimizer & AdamW \\
Learning rate schedule & Cosine annealing \\
Learning rate & $5.0 \times 10^{-3}$ \\
Weight decay & $3.4 \times 10^{-5}$ \\
Dropout rate & $4.1 \times 10^{-3}$ \\
Batch size & 512 \\
Maximum epochs & 4000 \\
Patience (early stopping) & 500 \\
\midrule
\multicolumn{2}{l}{\textit{Loss Function \& Physics Constraints}} \\
Primary loss & Mean squared error (MSE) \\
Huber loss parameter $\delta$ & 1.628 \\
Physics constraint types & Ratio consistency, Smoothness, Shape \\
Ratio constraint weight $\lambda_{\rm ratio}$ & $3.28 \times 10^{-2}$ \\
Smoothness constraint weight $\lambda_{\rm smooth}$ & $4.23 \times 10^{-2}$ \\
Shape constraint weight $\lambda_{\rm shape}$ & $7.75 \times 10^{-2}$ \\
Physics update frequency & Every 5 epochs \\
\midrule
\multicolumn{2}{l}{\textit{Performance Metrics (seed 450)}} \\
Validation MAE & 0.01463 \\
Overall yield error & 1.47\% \\
Training error & 1.08\% \\
Interpolation error & 1.78\% \\
Extrapolation error & 5.83\% \\
\bottomrule
\end{tabular}
\end{table}

\section{Boltzmann-Gibbs blast-wave fitting}
\label{app:2}
\begin{figure*}[h]
    \centering

    \begin{subfigure}{\textwidth}
        \centering
        \caption{Interpolation region}
        
        \includegraphics[width=\textwidth]{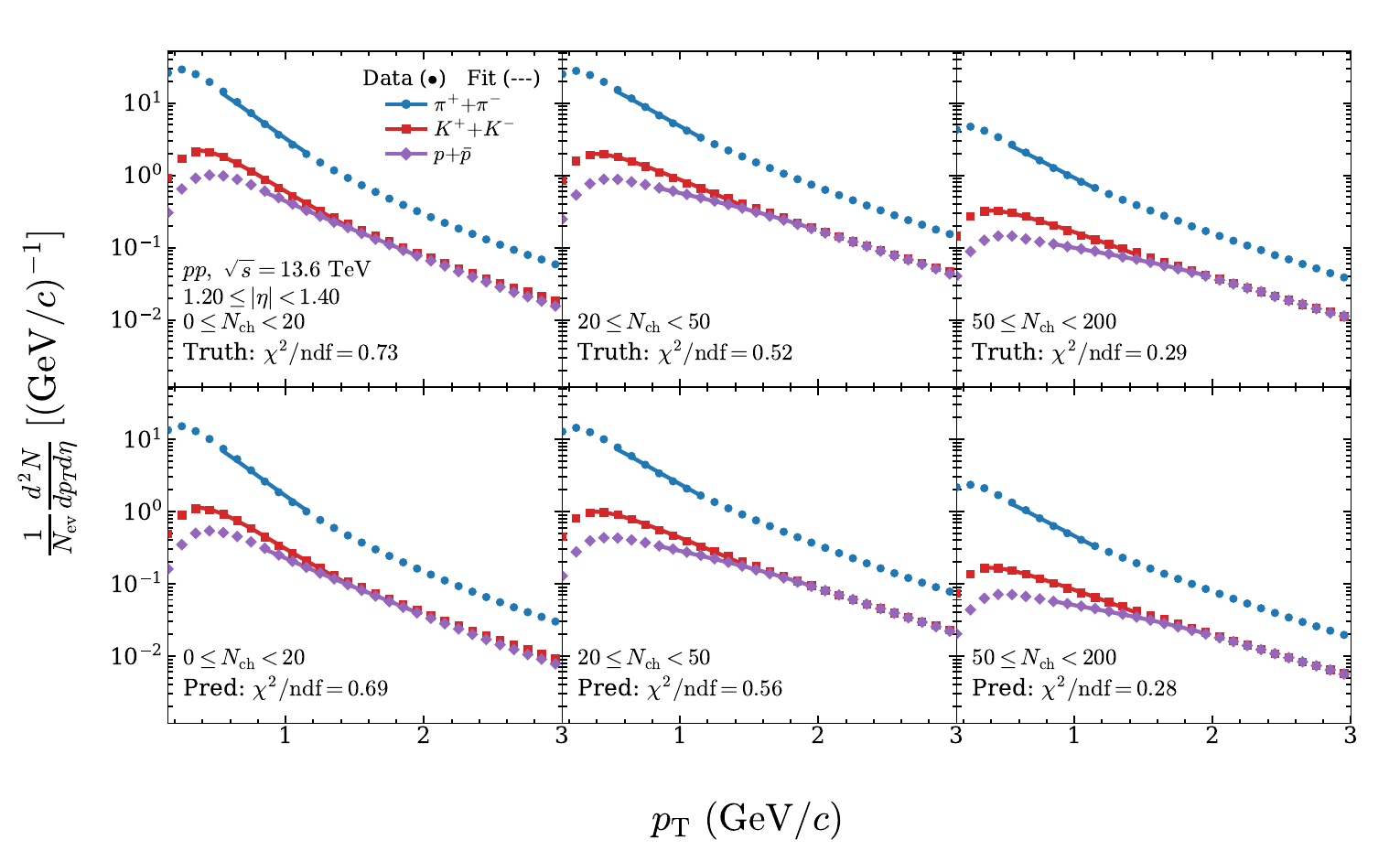}
        \label{fig:bmann_interp}
    \end{subfigure}

    \begin{subfigure}{\textwidth}
        \centering
        \caption{Extrapolation region}
        \includegraphics[width=\textwidth]{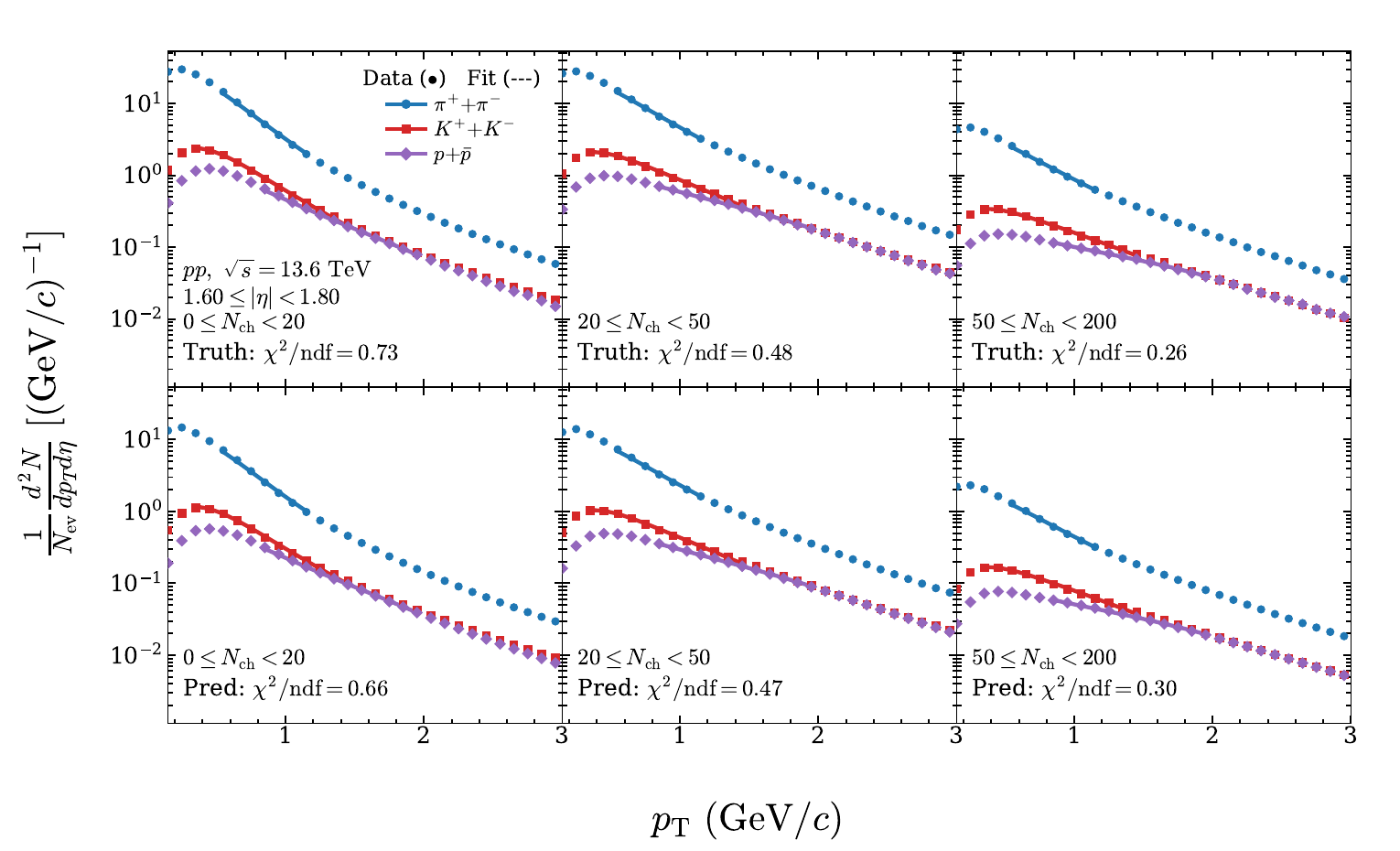}
        \label{fig:bmann_extrap}
    \end{subfigure}

    \caption{Simultaneous BGBW function fit to the identified charged particles' $\pT$ spectra in pp collisions at $\sqrt{s}=13.6$ TeV using PYTHIA8 and PINN predictions. Here, `Truth' refers to PYTHIA8, and `Pred' refers to the best PINN model predictions.}
    \label{fig:boltzmannfit}
\end{figure*}

In Fig.~\ref{fig:boltzmannfit} we present the Boltzmann-Gibbs blast-wave fits to the predicted and true spectra in the interpolation and extrapolation regions using Eq.~\eqref{eqn:bgbw}. The kinetic freeze-out parameters extracted from the fits for all $\eta$ ranges are listed in Table~\ref{tab:blastwave_all_eta}. As seen from the table, the model predictions in the training region agree very well with the parameters obtained from the PYTHIA8-generated spectrum. Moving toward the extrapolation region, a small deviation is observed for the $20\leq N_{\rm ch}<50$ class. Nevertheless, the overall trend remains consistent in both cases, indicating that the model successfully captures the spectral behavior in the regions where the fits are performed. The fitting was carried out up to $1.8 \leq |\eta| < 2.0$, for which the error remained within 4\%. Beyond this range, the error increases and produces unphysical results. Overall, the model learns correlations among the particle spectra, allowing it to predict them simultaneously while still producing physically meaningful results in regions not included in the training.

\begin{table*}
\centering
\caption{Blast-wave fit parameters comparing PYTHIA8 truth and PINN-predicted spectra across all $|\eta|$ and multiplicity bins. Fit windows: $\pT^{\pi}\in[0.5,1.0]$, $\pT^{K}\in[0.2,1.5]$, and $\pT^{p}\in[0.7,2.0]$~GeV/$c$.}
\label{tab:blastwave_all_eta}
\footnotesize
\setlength{\tabcolsep}{2.5pt}
\renewcommand{\arraystretch}{1.02}
\begin{tabular*}{\textwidth}{@{\extracolsep{\fill}}lc cc cc cc@{}}
\hline
\hline
 & & \multicolumn{2}{c}{$\langle\beta_T\rangle$} & \multicolumn{2}{c}{$T_{\mathrm{kin}}$\,(GeV)} & \multicolumn{2}{c}{$\chi^2/\mathrm{ndf}$} \\
\cline{3-4}\cline{5-6}\cline{7-8}
$|\eta|$ & $N_{\mathrm{ch}}$ & Truth & Pred. & Truth & Pred. & Truth & Pred. \\
\hline
\multirow{3}{*}{$ 0.0$--$0.2 $} & $0$--$20$ & $0.304\pm.037$ & $0.308\pm.034$ & $0.117\pm.009$ & $0.117\pm.008$ & 1.44 & 1.41 \\
 & $20$--$50$ & $0.468\pm.013$ & $0.467\pm.013$ & $0.102\pm.005$ & $0.102\pm.005$ & 0.95 & 0.95 \\
 & $50$--$200$ & $0.517\pm.012$ & $0.515\pm.012$ & $0.095\pm.005$ & $0.096\pm.005$ & 0.47 & 0.46 \\
\hline
\multirow{3}{*}{$ 0.2$--$0.4 $} & $0$--$20$ & $0.294\pm.048$ & $0.294\pm.053$ & $0.119\pm.012$ & $0.119\pm.013$ & 1.38 & 1.35 \\
 & $20$--$50$ & $0.461\pm.013$ & $0.464\pm.013$ & $0.103\pm.005$ & $0.102\pm.005$ & 0.93 & 0.94 \\
 & $50$--$200$ & $0.511\pm.012$ & $0.507\pm.012$ & $0.096\pm.005$ & $0.097\pm.006$ & 0.52 & 0.51 \\
\hline
\multirow{3}{*}{$ 0.4$--$0.6 $} & $0$--$20$ & $0.276\pm.047$ & $0.272\pm.037$ & $0.122\pm.011$ & $0.122\pm.009$ & 1.25 & 1.14 \\
 & $20$--$50$ & $0.448\pm.014$ & $0.451\pm.014$ & $0.105\pm.005$ & $0.103\pm.005$ & 0.88 & 0.83 \\
 & $50$--$200$ & $0.498\pm.012$ & $0.495\pm.012$ & $0.098\pm.005$ & $0.098\pm.005$ & 0.47 & 0.49 \\
\hline
\multirow{3}{*}{$ 0.6$--$0.8 $} & $0$--$20$ & $0.257\pm.037$ & $0.254\pm.031$ & $0.124\pm.009$ & $0.124\pm.007$ & 1.04 & 0.90 \\
 & $20$--$50$ & $0.434\pm.014$ & $0.432\pm.014$ & $0.106\pm.005$ & $0.106\pm.005$ & 0.77 & 0.83 \\
 & $50$--$200$ & $0.482\pm.013$ & $0.474\pm.014$ & $0.099\pm.005$ & $0.102\pm.006$ & 0.42 & 0.42 \\
\hline
\multirow{3}{*}{$ 0.8$--$1.0 $} & $0$--$20$ & $0.257\pm.030$ & $0.258\pm.027$ & $0.123\pm.007$ & $0.124\pm.006$ & 0.96 & 0.96 \\
 & $20$--$50$ & $0.406\pm.015$ & $0.408\pm.015$ & $0.110\pm.005$ & $0.110\pm.005$ & 0.66 & 0.70 \\
 & $50$--$200$ & $0.441\pm.015$ & $0.446\pm.013$ & $0.105\pm.006$ & $0.100\pm.005$ & 0.44 & 0.36 \\
\hline
\multirow{3}{*}{$ 1.0$--$1.2 $} & $0$--$20$ & $0.260\pm.029$ & $0.249\pm.029$ & $0.122\pm.007$ & $0.124\pm.007$ & 0.87 & 0.70 \\
 & $20$--$50$ & $0.385\pm.015$ & $0.388\pm.016$ & $0.113\pm.005$ & $0.112\pm.006$ & 0.57 & 0.61 \\
 & $50$--$200$ & $0.419\pm.014$ & $0.423\pm.014$ & $0.104\pm.006$ & $0.103\pm.006$ & 0.27 & 0.29 \\
\hline
\multirow{3}{*}{$ 1.2$--$1.4 $} & $0$--$20$ & $0.229\pm.006$ & $0.234\pm.010$ & $0.125\pm.003$ & $0.123\pm.003$ & 0.63 & 0.62 \\
 & $20$--$50$ & $0.375\pm.016$ & $0.377\pm.015$ & $0.114\pm.006$ & $0.112\pm.005$ & 0.52 & 0.56 \\
 & $50$--$200$ & $0.415\pm.015$ & $0.410\pm.015$ & $0.106\pm.006$ & $0.107\pm.006$ & 0.29 & 0.28 \\
\hline
\multirow{3}{*}{$ 1.4$--$1.6 $} & $0$--$20$ & $0.229\pm.007$ & $0.230\pm.005$ & $0.123\pm.004$ & $0.123\pm.003$ & 0.52 & 0.56 \\
 & $20$--$50$ & $0.362\pm.015$ & $0.356\pm.016$ & $0.115\pm.005$ & $0.117\pm.005$ & 0.48 & 0.51 \\
 & $50$--$200$ & $0.405\pm.015$ & $0.399\pm.016$ & $0.107\pm.006$ & $0.109\pm.006$ & 0.29 & 0.29 \\
\hline
\multirow{3}{*}{$ 1.6$--$1.8 $} & $0$--$20$ & $0.228\pm.002$ & $0.230\pm.005$ & $0.122\pm.003$ & $0.123\pm.003$ & 0.53 & 0.53 \\
 & $20$--$50$ & $0.356\pm.016$ & $0.347\pm.017$ & $0.117\pm.005$ & $0.122\pm.005$ & 0.49 & 0.48 \\
 & $50$--$200$ & $0.392\pm.016$ & $0.391\pm.016$ & $0.108\pm.006$ & $0.111\pm.006$ & 0.26 & 0.31 \\
\hline
\multirow{3}{*}{$ 1.8$--$2.0 $} & $0$--$20$ & $0.227\pm.002$ & $0.229\pm.002$ & $0.121\pm.003$ & $0.123\pm.003$ & 0.52 & 0.52 \\
 & $20$--$50$ & $0.339\pm.016$ & $0.333\pm.013$ & $0.117\pm.006$ & $0.121\pm.003$ & 0.47 & 0.47 \\
 & $50$--$200$ & $0.385\pm.015$ & $0.384\pm.017$ & $0.108\pm.006$ & $0.111\pm.006$ & 0.27 & 0.31 \\
\hline
\hline
\end{tabular*}
\end{table*}
\twocolumngrid


\clearpage

\begin{thebibliography}{99}

\bibitem{McLerran:1986zb}
L.~D.~McLerran,
\href{https://doi.org/10.1103/RevModPhys.58.1021}{Rev. Mod. Phys. \textbf{58}, 1021 (1986)}


\bibitem{Wong:1995jf}
C.~Y.~Wong,
\textit{Introduction to High-Energy Heavy-Ion Collisions}
\href{https://doi.org/10.1142/1128}{
(World Scientific, Singapore, 1994)}.

\bibitem{Gyulassy:2004zy}
M.~Gyulassy and L.~McLerran,
\href{https://doi.org/10.1016/j.nuclphysa.2004.10.034}{Nucl. Phys. A \textbf{750}, 30 (2005)}.

\bibitem{Shuryak:2014zxa}
E.~Shuryak,
\href{https://doi.org/10.1103/RevModPhys.89.035001}{Rev. Mod. Phys. \textbf{89}, 035001 (2017)}.

\bibitem{ALICE:2020nkc}
S.~Acharya \textit{et al.} (ALICE Collaboration),
\href{https://doi.org/10.1140/epjc/s10052-020-8125-1}{Eur. Phys. J. C \textbf{80}, 693 (2020) }.

\bibitem{Schnedermann:1993ws}
E.~Schnedermann, J.~Sollfrank and U.~W.~Heinz,
\href{https://doi.org/10.1103/PhysRevC.48.2462}{Phys. Rev. C \textbf{48}, 2462 (1993)}.

\bibitem{Prasad:2021bdq}
S.~Prasad, N.~Mallick, D.~Behera, R.~Sahoo and S.~Tripathy,
\href{https://doi.org/10.1038/s41598-022-07547-z}{Sci. Rep. \textbf{12}, 3917 (2022)}.

\bibitem{ALICE:2013mez}
B.~Abelev \textit{et al.} (ALICE Collaboration),
\href{https://doi.org/10.1103/PhysRevC.88.044910}{Phys. Rev. C \textbf{88}, 044910 (2013)}.

\bibitem{MenonKavumpadikkalRadhakrishnan:2023cik}
A.~Menon Kavumpadikkal Radhakrishnan, S.~Prasad, S.~Tripathy, N.~Mallick and R.~Sahoo,
\href{https://doi.org/10.1140/epjp/s13360-025-05996-9}{Eur. Phys. J. Plus \textbf{140}, 110 (2025)}.

\bibitem{ALICE:2008ngc}
K.~Aamodt \textit{et al.} (ALICE Collaboration),
\href{https://doi.org/10.1088/1748-0221/3/08/S08002}{JINST \textbf{3}, S08002 (2008)}

\bibitem{ALICE:2023udb}
S.~Acharya \textit{et al.} (ALICE Collaboration)
\href{https://doi.org/10.1088/1748-0221/19/05/P05062}{JINST \textbf{19}, P05062 (2024)}.

\bibitem{Sahoo:2025vip}
R.~Sahoo, K.~Goswami and S.~Prasad,
\href{https://doi.org/10.48550/arXiv.2509.00712}{arXiv:2509.00712} (Phys. Rev. D: In press).

\bibitem{Goswami:2024xrx}
K.~Goswami, S.~Prasad, N.~Mallick, R.~Sahoo and G.~B.~Mohanty,
\href{https://doi.org/10.1103/PhysRevD.110.034017}{Phys. Rev. D \textbf{110}, 034017 (2024)}.

\bibitem{Prasad:2023zdd}
S.~Prasad, N.~Mallick and R.~Sahoo,
\href{https://doi.org/10.1103/PhysRevD.109.014005}{Phys. Rev. D \textbf{109}, 014005 (2024)}.

\bibitem{Mallick:2021wop}
N.~Mallick, S.~Tripathy, A.~N.~Mishra, S.~Deb and R.~Sahoo,
\href{https://doi.org/10.1103/PhysRevD.103.094031}{Phys. Rev. D \textbf{103}, 094031 (2021)}.

\bibitem{Mallick:2023vgi}
N.~Mallick, S.~Prasad, A.~N.~Mishra, R.~Sahoo and G.~G.~Barnaf{\"o}ldi,
\href{https://doi.org/10.1103/PhysRevD.107.094001}{Phys. Rev. D \textbf{107}, 094001 (2023)}.

\bibitem{Mallick:2022alr}      
N.~Mallick, S.~Prasad, A.~N.~Mishra, R.~Sahoo and G.~G.~Barnaf{\"o}ldi,
\href{https://doi.org/10.1103/PhysRevD.105.114022}{Phys. Rev. D \textbf{105}, 114022 (2022)}.

\bibitem{Tiwari:2026vef}
R.~K.~Tiwari, K.~Goswami, S.~Prasad, C.~R.~Singh, R.~Sahoo and M.~Y.~Jamal,
\href{https://doi.org/10.48550/arXiv.2604.05858}{arXiv:2604.05858}.

\bibitem{mage_ref}
 R.~ Gupta, K.~Goswami,  S.~Prasad, and R.~Sahoo, MAGE, paper in preparation.

\bibitem{Bierlich:2026syi}
C.~Bierlich, L.~L{\"o}nnblad and T.~Sj{\"o}strand,
\href{https://doi.org/10.48550/arXiv.2603.01744}{arXiv:2603.01744.}

\bibitem{Sjostrand:2014zea}
T.~Sj{\"o}strand, S.~Ask, J.~R.~Christiansen, R.~Corke, N.~Desai, P.~Ilten, S.~Mrenna, S.~Prestel, C.~O.~Rasmussen and P.~Z.~Skands,
\href{https://doi.org/10.1016/j.cpc.2015.01.024}{Comput. Phys. Commun. \textbf{191}, 159 (2015)}

\bibitem{Sjostrand:2006za}
T.~Sjostrand, S.~Mrenna and P.~Z.~Skands,
\href{https://doi.org/10.1088/1126-6708/2006/05/026}{JHEP \textbf{05}, 026 (2006)}.

\bibitem{Prasad:2025yfj}
S.~Prasad, S.~Tripathy, B.~Sahoo and R.~Sahoo,
\href{https://doi.org/10.1016/j.physrep.2026.04.001}{Phys. Rept. \textbf{1181}, 1 (2026)}.

\bibitem{Prasad:2024gqq}
S.~Prasad, B.~Sahoo, S.~Tripathy, N.~Mallick and R.~Sahoo,
\href{https://doi.org/10.1103/PhysRevC.111.044902}{Phys. Rev. C \textbf{111}, 044902 (2025)}.

\bibitem{ALICE:2023bga}
S.~Acharya \textit{et al.} (ALICE Collaboration),
\href{https://doi.org/10.1007/JHEP05(2024)184}{JHEP \textbf{05}, 184 (2024)}.

\bibitem{ALICE:2015ikl}
J.~Adam \textit{et al.} (ALICE Collaboration),
\href{https://doi.org/10.1007/JHEP09(2015)148}{JHEP \textbf{09}, 148 (2015)}.

\bibitem{Thakur:2017kpv}
D.~Thakur, S.~De, R.~Sahoo and S.~Dansana,
\href{https://doi.org/10.1103/PhysRevD.97.094002}{Phys. Rev. D \textbf{97}, 094002 (2018)}.

\bibitem{Deb:2018qsl}
S.~Deb, D.~Thakur, S.~De and R.~Sahoo,
\href{https://doi.org/10.1140/epja/s10050-020-00138-4}{Eur. Phys. J. A \textbf{56}, 134 (2020)}.

\bibitem{Brun:1997pa}
R.~Brun and F.~Rademakers,
\href{https://doi.org/10.1016/s0168-9002(97)00048-x}{Nucl. Instrum. Meth. A \textbf{389}, 81 (1997)}.

\bibitem{Skands:2014pea}
P.~Skands, S.~Carrazza and J.~Rojo,
\href{https://doi.org/10.1140/epjc/s10052-014-3024-y}{Eur. Phys. J. C \textbf{74}, 3024 (2014)}.

\bibitem{ml:xgboost}               
T.~Chen and C.~Guestrin,
in \textit{Proceedings of the 22nd ACM SIGKDD International Conference on
Knowledge Discovery and Data Mining}
(ACM, New York, 2016), pp.~785,\href{https://doi.org/10.1145/2939672.2939785}{10.1145/2939672.2939785}.

\bibitem{ml:2017lightgbm}
G.~Ke, Q.~Meng, T.~Finley, T.~Wang, W.~Chen, W.~Ma, Q.~Ye and T.~Y.~Liu,
in \textit{Advances in Neural Information Processing Systems 30 (NIPS 2017)}
(Curran Associates, Inc., 2017), pp.~3149,\url{https://dl.acm.org/doi/10.5555/3294996.3295074}.

\bibitem{Shokr:2021ouh}
E.~Shokr, A.~De Roeck and M.~A.~Mahmoud,
\href{https://doi.org/10.1038/s41598-022-11618-6}{Sci. Rep. \textbf{12}, 8449 (2022)}.

\bibitem{ml:xgboost_docs}
Available Online: 
\url{https://xgboost.readthedocs.io/en/stable/}.

\bibitem{ml:lightgbm_docs}     
Available Online: 
\url{https://lightgbm.readthedocs.io/en/latest/}.

\bibitem{Karniadakis:2021}
G.~E.~Karniadakis, I.~G.~Kevrekidis, L.~Lu, P.~Perdikaris, S.~Wang and L.~Yang,
\href{https://doi.org/10.1038/s42254-021-00314-5}
{Nat. Rev. Phys. \textbf{3}, 422 (2021)}.

\bibitem{Raissi:2019pinn}
M. Raissi, P. Perdikaris, and G. E. Karniadakis,
\href{https://doi.org/10.1016/j.jcp.2018.10.045}
{J. Comput. Phys. \textbf{378}, 686 (2019)}.

\bibitem{Bezdek2003}         
J.~C.~Bezdek and R.~J.~Hathaway,
\href{https://doi.org/10.1007/3-540-45631-7_39}
{in \textit{Advances in Soft Computing}, Vol. \textbf{2275}
(Springer, Berlin, Heidelberg, 2002), pp. 288}.

\bibitem{Watanabe:2023tpe}
S.~Watanabe,
\href{https://doi.org/10.48550/arXiv.2304.11127}{arXiv:2304.11127}.

\bibitem{Akiba:2019optuna}       
T.~Akiba, S.~Sano, T.~Yanase, T.~Ohta and M.~Koyama,
\href{https://doi.org/10.48550/arXiv.1907.10902}{arXiv:1907.10902}.

\bibitem{Jiang:2024} 
J.~Jiang and A.~Mian,
\href{https://doi.org/10.48550/arXiv.2409.00584}{arXiv:2409.00584}.

\bibitem{Stefan:2018} 
S.~Falkner, A.~Klein and F.~Hutter,
\href{https://doi.org/10.48550/arXiv.1807.01774}{arXiv:1807.01774}.

\bibitem{Pascanu:2012}
R.~Pascanu, T.~Mikolov and Y.~Bengio,
\href{https://doi.org/10.48550/arXiv.1211.5063}
{arXiv:1211.5063}.

\bibitem{Siemens:1978pb}
P.~J.~Siemens and J.~O.~Rasmussen,
\href{https://doi.org/10.1103/PhysRevLett.42.880}{Phys. Rev. Lett. \textbf{42}, 880 (1979)}.

\bibitem{OrtizVelasquez:2013ofg}
A.~Ortiz Velasquez, P.~Christiansen, E.~Cuautle Flores, I.~Maldonado Cervantes and G.~Pai{\'c},
\href{https://doi.org/10.1103/PhysRevLett.111.042001}{Phys. Rev. Lett. \textbf{111}, 042001 (2013)}.

\bibitem{Retiere:2003kf}
F.~Retiere and M.~A.~Lisa,
\href{https://doi.org/10.1103/PhysRevC.70.044907}{Phys. Rev. C \textbf{70}, 044907 (2004)}.

\bibitem{ALICE:2016fzo}
J.~Adam \textit{et al.} (ALICE Collaboration),
\href{https://doi.org/10.1038/nphys4111}{Nature Phys. \textbf{13}, 535 (2017)}.

\bibitem{ALICE:2018pal}
S.~Acharya \textit{et al.} (ALICE Collaboration),
\href{https://doi.org/10.1103/PhysRevC.99.024906}{Phys. Rev. C \textbf{99}, 024906 (2019)}.

\bibitem{ALICE:2013rdo}
B.~B.~Abelev \textit{et al.} (ALICE Collaboration),
\href{https://doi.org/10.1016/j.physletb.2013.10.054}{Phys. Lett. B \textbf{727}, 371 (2013)}.

\bibitem{Braun-Munzinger:1994ewq}
P.~Braun-Munzinger, J.~Stachel, J.~P.~Wessels and N.~Xu,
\href{https://doi.org/10.1016/0370-2693(94)01534-J}{Phys. Lett. B \textbf{344}, 43 (1995)}.

\bibitem{Huovinen:2001cy}
P.~Huovinen, P.~F.~Kolb, U.~W.~Heinz, P.~V.~Ruuskanen and S.~A.~Voloshin,
\href{https://doi.org/10.1016/S0370-2693(01)00219-2}{Phys. Lett. B \textbf{503}, 58 (2001)}.

\bibitem{ALICE:2022wpn}
S.~Acharya \textit{et al.} (ALICE Collaboration),
\href{https://doi.org/10.1140/epjc/s10052-024-12935-y}{Eur. Phys. J. C \textbf{84}, 813 (2024)}.

\bibitem{Ortiz:2015ttf}
A.~Ortiz, G.~Pai{\'c} and E.~Cuautle,
\href{https://doi.org/10.1016/j.nuclphysa.2015.05.010}{Nucl. Phys. A \textbf{941}, 78 (2015)}.

\bibitem{BRAHMS:2016klg}
I.~C.~Arsene \textit{et al.} (BRAHMS Collaboration),
\href{https://doi.org/10.1103/PhysRevC.94.014907}{Phys. Rev. C \textbf{94}, 014907 (2016)}.

\bibitem{Tang:2011xq}
Z.~Tang, L.~Yi, L.~Ruan, M.~Shao, H.~Chen, C.~Li, B.~Mohanty, P.~Sorensen, A.~Tang and Z.~Xu,
Chin. \href{https://doi.org/10.1088/0256-307X/30/3/031201}{Phys. Lett. \textbf{30}, 031201 (2013)}.

\bibitem{PHENIX:2003wtu}
K.~Adcox \textit{et al.} (PHENIX Collaboration),
\href{https://doi.org/10.1103/PhysRevC.69.024904}{Phys. Rev. C \textbf{69}, 024904 (2004)}.

\end{thebibliography}
\end{document}